\documentclass[%
 reprint,
 superscriptaddress,
 showpacs,preprintnumbers,
 amsmath,amssymb,
 aps,
 prb,
]{revtex4-1}
\usepackage{graphicx}
\usepackage{dcolumn}% Align table columns on decimal point
\usepackage{bm}% bold math
\usepackage[usenames]{color}
\begin{document}

\title[Formation of Co/Ge intermixing layers on Ge(111)2$\times$1 surfaces]{Formation of Co/Ge intermixing layers after Co deposition on Ge(111)2$\times$1 surfaces}

%(c) Koen. Some suggestions/ideas for a title:
%-Formation of Co/Ge (sub)surface intermixing layers by Co atom deposition on Ge(111)2x1: A LT STM study
%-Formation of a novel surface reconstruction by Co atom deposition on Ge(111)2x1: A LT STM study
%-STM study of a novel surface reconstruction formed after Co atom deposition on Ge(111)2x1
%-STM study of (sub)surface intermixing layers formed after Co atom deposition on Ge(111)2x1

\author{D.~A.~Muzychenko}
\email{mda@spmlab.ru}
\affiliation{Faculty of Physics, Moscow State University, 119991 Moscow, Russia}
\author{K.~Schouteden}
\email{Koen.Schouteden@fys.kuleuven.be}
\affiliation{Laboratory of Solid-State Physics and Magnetism, KULeuven, BE-3001 Leuven, Belgium}
\author{V.~I.~Panov}
\affiliation{Faculty of Physics, Moscow State University, 119991 Moscow, Russia}
\author{C.~Van~Haesendonck}
\affiliation{Laboratory of Solid-State Physics and Magnetism, KULeuven, BE-3001 Leuven, Belgium}

%========== Authors for IOP Style ==================
%\author{D.~A.~Muzychenko$^1$, K.~Schouteden$^2$, V.~I.~Panov$^1$ and C.~Van~Haesendonck$^2$}
%\address{$^1$ Faculty of Physics, Moscow State University, 119991 Moscow, Russia}
%\address{$^2$ Laboratory of Solid-State Physics and Magnetism, BE-3001 Leuven, Belgium}
%\ead{mda@spmlab.ru}
%==================================================

\date{\today}

\begin{abstract}
The formation of a novel surface reconstruction upon Co deposition on freshly cleaved Ge(111)2$\times1$ surfaces is studied by means of scanning tunneling microscopy (STM) at low temperature. The deposited Co atoms are immobile at substrate temperatures of $4.5 \, {\rm K}$, while they can diffuse along the upper $\pi$-bonded chains at a temperature of $80 \, {\rm K}$ and higher. This mobility results in accumulation of Co atoms at atomic steps, at domain boundaries as well as on atomically flat Ge terraces at, e.g., vacancies or adatoms, where reconstructed Co/Ge intermixing layers are formed. Voltage dependent STM images reveal that the newly reconstructed surface locally exhibits a highly ordered atomic structure, having the same periodicity as that of the initial 2$\times$1 reconstruction. In addition, it shows a double periodicity along the $[2\overline{11}]$ direction, which can be related to the modified electronic properties of the $\pi$-bonded chains.
\end{abstract}

%\keywords{STM, STS, Low Temperature, Semiconductor, Germanium, Ge(111), 2x1, Surface, Surface Reconstruction, Co, Cobalt, Single Atom, Atom Diffusion, Atom Migration, Atom Embedding, Zero-dimensional, 0D, Intermixing Layer, Alloys, Germanides, Epitaxial Layer, Nanofilms, Nanostructures, Nanoelectronics, One-dimensional, 1D, Two-dimensional, 2D, Nanowires, Quantum Dots}

\maketitle

%%%%%%%%%%%%%%%%%%%%%%%%%%%%%%%%%%%%%%%%%%%%%%%%%%%%%%%%%%%%%%%%%%%%%%%%%%%%
%%%%%%%%%%%%%%%%%%%%%%%%%%%%%%%%%%%%%%%%%%%%%%%%%%%%%%%%%%%%%%%%%%%%%%%%%%%%
%%%%%%%%%%%%%%%%%%            Introduction                  %%%%%%%%%%%%%%%%
%%%%%%%%%%%%%%%%%%%%%%%%%%%%%%%%%%%%%%%%%%%%%%%%%%%%%%%%%%%%%%%%%%%%%%%%%%%%
%%%%%%%%%%%%%%%%%%%%%%%%%%%%%%%%%%%%%%%%%%%%%%%%%%%%%%%%%%%%%%%%%%%%%%%%%%%%
\section{Introduction}
During the past few years, many advance has been made by the scientific community in the nanoelectronic industry, in overcoming the difficulties that are encountered in the undiminished miniaturization process of electronic devices.~\cite{Haider_PRL_09,Tan_NanoLett_09,Parks_Science_10} However, in order to further continue this miniaturization trend, new materials with better electronic properties than that of Si are required. Among many materials, Ge is today considered as a promising alternative~\cite{Chui_APL_03,Bracht_PRL_09,Wundisch_APL_09} because of its high carrier mobility and its compatibility with current Si based technology.~\cite{Sze_SolStEl_68} In this view, metal/Ge metal/high-\emph{k} and Ge-based alloys have received considerable interest in the past few years because they exhibit a Schottky barrier at the interface.~\cite{Rowe_PRL_75,Chang_PRL_89,Lee_MatTod_06,Koon-Yiu_PRL_07}

In analogy with the current Si based technology, where metal silicides (formed by thermal reaction of a metal layer with the Si substrate) are used to contact the source, drain, and gate regions of the transistors,~\cite{Lee_MatTod_06} metal germanides can be used for the production of self-aligned contacts.~\cite{Zhu_IEEE_05,Park_JES_07,Brunco_JES_08} However, only a fraction of these germanides has potential to be used as electrical contacts, since the electrical resistance of the germanide phase must be low and it must have a high thermal stability. A detailed overview of the subsequent germanide phases that are formed during thermal reaction of Ge with various metals is provided by Gaudet \emph{et al.}~\cite{Gaudet_JVSTA_06} Three materials of this list fulfil the required criteria: CoGe$_2$, NiGe and PdGe. The phase evolution of Co- and Ni-germanides as a function of temperature has previously been been studied in great detail,~\cite{Ashburn_JAP_93,Gaudet_JAP_06,Opsomer_APL_07} as well as their Schottky diode behavior.~\cite{Chi_JAP_05,Simoen_APL_06,Simoen_JAP_08} NiGe and PdGe germanides due to their low formation temperature and low resistivity are considered as main candidates as contacting materials for Ge channel devices.~\cite{Kittl_MSEB_08}

The growth of thin (magnetic) films, clusters or nanowires of these materials on semiconducting surfaces may be exploited to design novel spin-based electronic devices.~\cite{Wolf_Sci_01,Zutic_RMP_04} Spintronics requires the combined use of semiconducting and ferromagnetic materials in order to control the degree of electron spin.~\cite{Ryan_PRB_04} The combination of Co and Ge is considered as an important candidate for this purpose. Although the Co/Ge system has already been studied intensively~\cite{Ashburn_JAP_93,Mello_JAP_97,Dhar_TSF_98,Goldfarb_JMR_00,Tsay_SS_04,Sun_APL_05,Chang_JAP_06,Sell_EPJD_07,Park_JECS_09,Keyser_JECS_10}, the initial adsorption stage of Co atoms on Ge surfaces has not yet been investigated. Deeper understanding of the formation process of the Co/Ge interface upon Co adsorption is however of great technological and fundamental interest.

In this paper, we present a low-temperature (LT) scanning tunneling microscopy (STM) study of the initial growth stage of submonolayer coverages of Co deposited on Ge(111) surfaces. Clean Ge(111) surfaces are obtained by cleavage under ultra-high vacuum (UHV) conditions at room temperature (RT) and reveal the typical 2$\times$1 reconstruction, consisting of $\pi$-bonded chains of Ge atoms running in the $[01\overline{1}]$ direction.~\cite{Pandey_PRL_81,Northrup_PRB_83} Previously we found by combined STM experiments and density functional theory calculations that the deposited Co atoms are embedded non-invasively into the Ge(111)2$\times$1 surface. Here, we show that the embedded Co atoms can diffuse along the upper $\pi$-bonded chains along the $[01\overline{1}]$ direction and that they accumulate at atomic steps, domain boundaries, vacancies and adatoms to form Co/Ge intermixing layers that are accompanied by a novel Ge-Co reconstruction.

%%%%%%%%%%%%%%%%%%%%%%%%%%%%%%%%%%%%%%%%%%%%%%%%%%%%%%%%%%%%%%%%%%%%%%%%%%%%
%%%%%%%%%%%%%%%%%%%%%%%%%%%%%%%%%%%%%%%%%%%%%%%%%%%%%%%%%%%%%%%%%%%%%%%%%%%%
%%%%%%%%%%%%%             Experimental details               %%%%%%%%%%%%%%%
%%%%%%%%%%%%%%%%%%%%%%%%%%%%%%%%%%%%%%%%%%%%%%%%%%%%%%%%%%%%%%%%%%%%%%%%%%%%
%%%%%%%%%%%%%%%%%%%%%%%%%%%%%%%%%%%%%%%%%%%%%%%%%%%%%%%%%%%%%%%%%%%%%%%%%%%%

\section{Experimental details}

STM measurements were performed with a LT UHV setup (Omicron Nanotechnology), consisting of a room-temperature sample preparation chamber and a LT STM measurement chamber. The operating pressure in the chambers is about $5 \times 10^{-11}\, {\rm mbar}$ and $4 \times 10^{-12} \, {\rm mbar}$, respectively. In order to optimize measurement stability, the LT UHV setup is decoupled from the building by a specially designed vibration isolation floor. Electrochemically etched tungsten tips were used in all experiments. The tips were cleaned \emph{in situ} by repeated flashing above $1800 \, {\rm K}$ in order to remove the surface oxide layer and additional contamination. The tip quality was routinely checked by acquiring atomic resolution images of the ``herringbone'' reconstruction of the Au(111) surface.~\cite{Barth_JV_90,Schouteden_Nanotech_09} All here reported STM experiments are performed at liquid helium temperature ($T_{\rm sample} \, \simeq 4.5 \, {\rm K}$). STM topographic imaging was performed in constant current mode. Everywhere in the text the tunneling bias voltage $V_{\rm t}$ refers to the sample voltage, while the STM tip is virtually grounded. Image processing was performed by Nanotec WSxM.~\cite{WXsM}

The investigated Ge samples are doped with Ga at a doping level of $n_{Ga} \, = \, 1$ to $2\times 10^{16} \, {\rm cm}^{-3}$, resulting in $p$-type bulk conductivity with resistivity of $\rho_{\rm bulk} \simeq 0.2 \, {\rm \Omega~cm}$. $4 \times 1.5 \times 0.8 \, {\rm mm}^3$ Ge bars with their long axis aligned with the (111) direction were cleaved {\it in situ} in the preparation chamber. The freshly cleaved samples were transferred within about 5 minutes to the STM measurement chamber. The cleaved Ge(111)2$\times$1 surfaces were observed to retain their cleanliness for 5 to 7 days in the LT STM measurement chamber.

The experiments consist of three stages. First, the clean 2$\times$1 reconstructed surface of the freshly cleaved Ge single crystal is characterized in detail by STM measurements.~\cite{Muzychenko_PRB_10} Second, $0.02$ to $0.04$ monolayers (MLs) of Co atoms are deposited on the cold Ge surface ($T_{\rm sample} \, \leq 80 \, {\rm K}$) by means of an electron-beam evaporator at a rate of $0.007 \pm 0.001$ MLs per second. A high purity Co (99.9996\%) rod is used for Co atom evaporation. Third, the Co/Ge(111)2$\times$1 surface was investigated by LT STM measurements.

%%%%%%%%%%%%%%%%%%%%%%%%%%%%%%%%%%%%%%%%%%%%%%%%%%%%%%%%%%%%%%%%%%%%%%%%%%%%
%%%%%%%%%%%%%%%%%%%%%%%%%%%%%%%%%%%%%%%%%%%%%%%%%%%%%%%%%%%%%%%%%%%%%%%%%%%%
%%%%%%%%%%%%%            Results and Discussion              %%%%%%%%%%%%%%%
%%%%%%%%%%%%%%%%%%%%%%%%%%%%%%%%%%%%%%%%%%%%%%%%%%%%%%%%%%%%%%%%%%%%%%%%%%%%
%%%%%%%%%%%%%%%%%%%%%%%%%%%%%%%%%%%%%%%%%%%%%%%%%%%%%%%%%%%%%%%%%%%%%%%%%%%%
\section{Results and Discussion}

%--------------------------------------------------------------------------%
\subsection{The freshly cleaved Ge(111)2$\times$1 surface}
\label{subsect:Ge(111)2x1}
%--------------------------------------------------------------------------%
Prior to investigation of the newly formed Co-related structures, a careful characterization of the Ge(111)2$\times$1 surface is obviously required. For this purpose we have performed STM measurements on 7 freshly cleaved Ge crystals. After analysis of up to $1.9 \, {\rm \mu m}^{2}$  atomic resolution STM topography images of clean Ge(111)2$\times$1 surfaces, we can state that the following surface structures exist after cleavage: (i) atomically flat terraces that exhibit the 2$\times$1 surface reconstruction; (ii) mono atomic steps (MASs) of type-A and type-B; (iii) domain boundaries (DBs) of type-A and type-B; (iv) Ge adatoms (and vacancies); (v) Ga impurities and impurity/adatom complexes.

%=======================================================%
%=======    Fig 1  " Cleaved_Ge2x1_surface "   =========%
%=======================================================%
\begin{figure}
\begin{center}\includegraphics[scale=1.0]{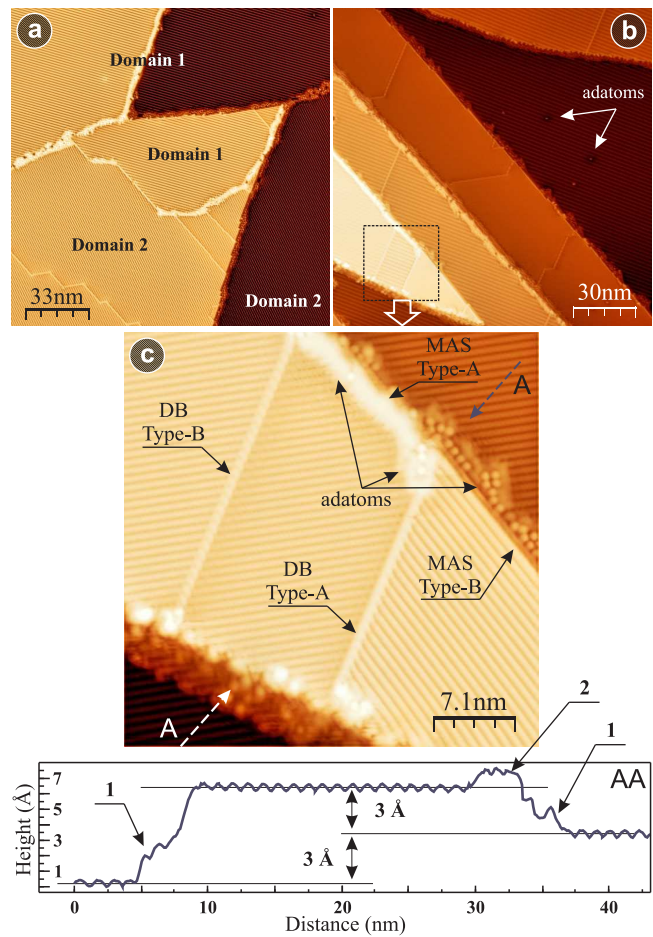}\end{center}
\caption{(Color online) (a) and (b) Typical large-scale STM images of the freshly cleaved Ge(111)2$\times$1 surface ($V_{\rm t} = +1.0 \, {\rm V}$,  $I_{\rm t} = 20 \, {\rm pA}$). (c) STM image of the area enclosed by the dashed square in (b). (bottom) The height profile is taken along the line in between the two dashed arrows with label A.}
\label{Cleaved_Ge2x1_surface}
\end{figure}
%=======================================================%
%=======================================================%
%=======================================================%

In Fig.~\ref{Cleaved_Ge2x1_surface} we present two typical large-scale STM images (Figs.~\ref{Cleaved_Ge2x1_surface}~(a) and (b)) and a high-resolution STM image of the clean Ge(111)2$\times$1 surface [Fig.~\ref{Cleaved_Ge2x1_surface}~(c)]. Large atomically flat terraces up to $10^{5} {\rm nm}^{2}$ can be easily retrieved, which are separated from each other by MASs. The terraces reveal the typical 2$\times$1 reconstruction, consisting of $\pi$-bonded chains of Ge atoms~\cite{Pandey_PRL_81,Northrup_PRB_83,Feenstra_SS_91} running in the $[01\overline{1}]$ direction. It is known that only the upper chains are visualized by STM.~\cite{Feenstra_PRB_01} The surface unite cell contains two atoms, both having one dangling bond that is responsible for $\pi$-bonding along the surface upper chains. In the original Pandey geometry model,~\cite{Pandey_PRL_81} two of the atoms of the 7-members Ge ring are considered to be physically at the same height at the Ge(111) surface, forming the zigzag chain along the $[01\overline{1}]$ direction. Due to buckling, however, one of these two atoms (referred to as the \emph{up}-atom) is shifted somewhat out of the surface while the other (referred to as the \emph{down}-atom) is shifted into the surface.~\cite{Takeuchi_PRB_91} The occupied surface states are mainly localized to the \emph{upper}-atom, while the empty surface states are mainly localized to the \emph{down}-atom of the $\pi$-bonded chain. Consequently, the bonding surface states band $\pi_{\rm VB}$ derived from the \emph{up}-atom orbital is filled, while the anti-bonding surface states band $\pi_{\rm CB}^*$ derived from the \emph{down}-atom orbital is empty. For low doping concentrations, the anti-bonding surface states band $\pi_{\rm CB}^*$ remains unoccupied and two surface band gaps can be discerned in the STS spectra: (i) a wide gap corresponding to the bulk band gap, projected onto the Ge(111)2$\times$1 surface ($E_{\rm gs} = 0.54 \, {\rm eV}$); (ii) a narrow gap corresponding to the band gap between the top of the (occupied) bulk valence band \emph{VB} and the bottom of the (unoccupied) surface states band $\pi_{CB}^*$ ($E_{\rm gvs} = 0.19 \, {\rm eV}$). For more details on the relation between the surface and bulk bands for low doped Ge(111)2$\times$1 surfaces we refer to Refs.~\onlinecite{Muzychenko_PRB_10,Muzychenko_Submitted_11}.

It can be observed in Fig.~\ref{Cleaved_Ge2x1_surface} that the Ge(111)2$\times$1 surface consists of different types of domains with slightly different atomic arrangement.~\cite{Einaga_PRB_98} This is related to the threefold rotational symmetry of the surface. The domains are separated by two different types of DBs. In the first type of DB, referred to as type-A DB following the terminology used in Ref.~\onlinecite{Einaga_PRB_98}, the chains at the opposite sides of the DB are rotated by an angle of $\pi/3$ [see Fig.~\ref{Cleaved_Ge2x1_surface}~(c)]. The second type DB, the so called anti-phase DB or type-B DB~\cite{Einaga_PRB_98} is formed due to a shift of the $\pi$-bonded chains in the $[2\overline{11}]$ direction by a half-unit cell [see Fig.~\ref{Cleaved_Ge2x1_surface}(c)]. We found that most DB are of type-B~\cite{Muzychenko_PRB_10} and that the type-A DB often exhibits (local) disorder.

In addition, two types of MASs were observed [see Fig.~\ref{Cleaved_Ge2x1_surface}~(a) and (c)]. One type of MASs, hereafter referred to as type-A, is oblique to the $\pi$-bonded chains on the terrace. The second type of MASs, hereafter referred to as type-B, is parallel to the $\pi$-bonded chains. It is observed that Ge adatoms are often present at these MASs,~\cite{Feenstra_SS_91} both on the upper and lower terrace, except at the top terrace of type-B MAS [see Fig.~\ref{Cleaved_Ge2x1_surface}~(c)]. Here, the surface is either locally distorted or a 2$\times$2, 2$\times$4 or \emph{c}2$\times$8 surface reconstruction is formed [indicated in Fig.~\ref{Cleaved_Ge2x1_surface}~(c)]. In the STM images, Ge adatoms at MASs appear as bright protrusions on top of the 2$\times$1 surface reconstruction in the investigate voltage range $V_{\rm t} = \pm 2.5 \, {\rm V}$, as can be seen in the height profile in Figure~\ref{Cleaved_Ge2x1_surface}~(c), where the Ge adatoms at the lower and at the upper terraces near the type-A MAS are indicated by labels (1) and (2). These Ge surface adatoms very likely are created during cleavage at RT, after which they can migrate along $\pi$-bonded chains to the MASs. Furthermore, (individual) Ge adatoms could be frequently retrieved at atomically flat Ge(111)2$\times$1 terrace as well, i.e., above a charged subsurface Ga impurity [see Fig.~\ref{Cleaved_Ge2x1_surface}~(b)]. These adatoms are well separated and correlate with the low doping level of our Ge samples.

%--------------------------------------------------------------------------%
\subsection{Co deposition on cold Ge(111)2$\times$1 surfaces}
\label{subsect:Co_deposition}
\begin{figure}
\begin{center}\includegraphics[scale=1.0]{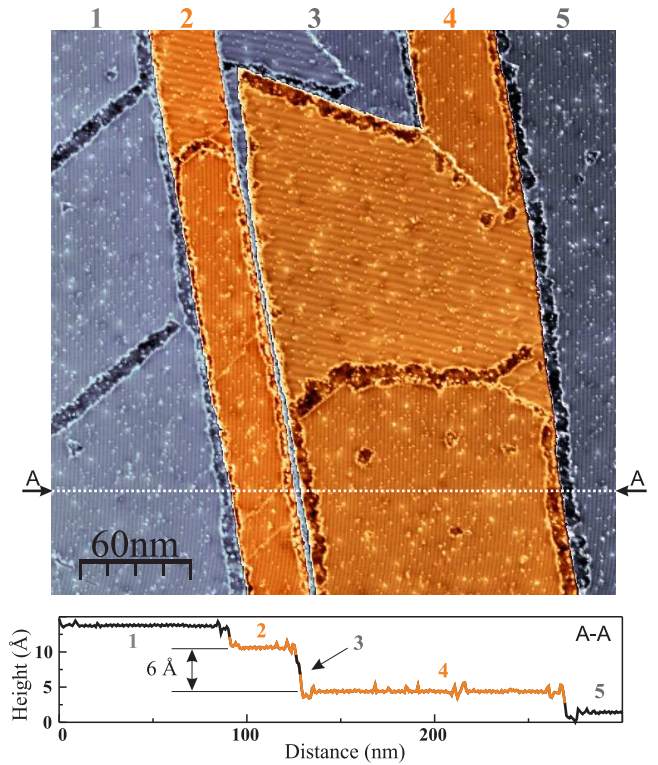}\end{center}
\caption{(Color online) (top) STM image of five Ge terraces after deposition of $0.032 \, {\rm MLs}$ of Co atoms on a cold ($T_s \leq 80 \, {\rm K}$) Ge(111)2$\times$1 surface ($V_{\rm t} = +1.0 \, {\rm V}$ and $I_{\rm t} = 15 \, {\rm pA}$). (bottom) Height profile taken along the white dotted line.}
\label{Global_changes}
\end{figure}

%--------------------------------------------------------------------------%
%======  Co deposition results    ======%
Four freshly cleaved Ge crystals have been investigated after Co atom deposition and a surface area of up to $2.4 \, {\rm \mu m}^{2}$ has been visualized by STM with atomic resolution. Figure~\ref{Global_changes} presents a typical large scale STM image of the Ge(111)2$\times$1 surface after deposition of $0.032$ MLs of Co atoms on a cold Ge(111)2$\times$1 surface ($T_{\rm sample} \, \leq 80 \, {\rm K}$). Apart from Co atoms that are uniformly spread across the atomically flat Ge terraces, it can be seen that novel reconstructed structures are formed, e.g., at MASs and DBs. At this voltage, these structures appear somewhat lower than the atomically flat terraces in the STM images and cover about $14.3$\% of the entire surface area and are referred to as Co/Ge intermixing layers (ILs) hereafter.

%=======================================================%
%=========     Fig 2  " Global_changes "     ===========%
%=======================================================%
%=======================================================%
%=======================================================%
%=======================================================%

\begin{figure}
\begin{center}\includegraphics{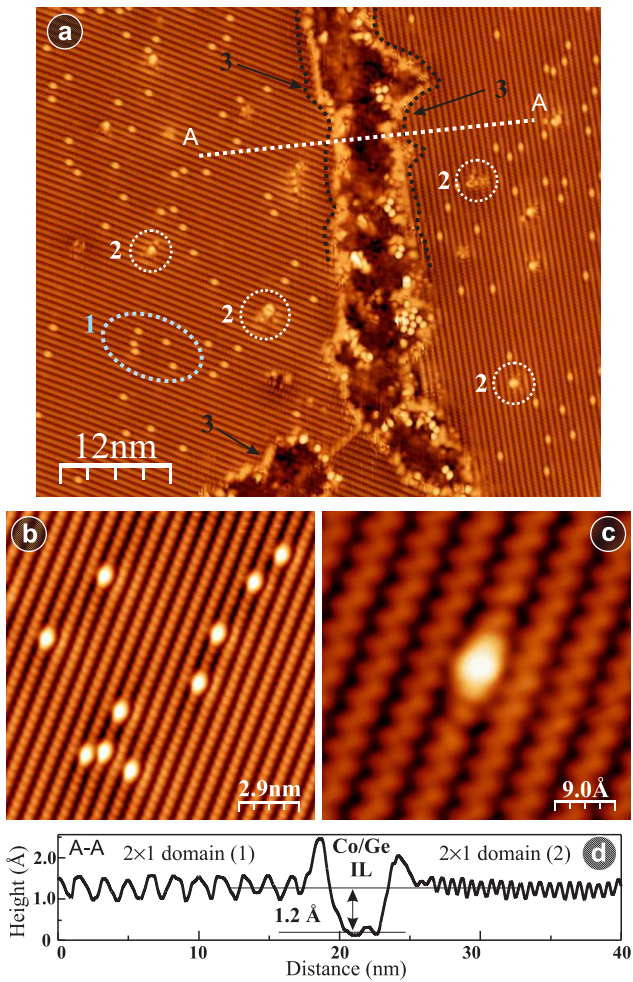}\end{center}
\caption{(Color online) (a) STM image of the Ge(111)2$\times$1 surface after Co deposition ($V_{\rm t} = +1.0 \, {\rm V}$, $I_{\rm t} = 15 \, {\rm pA}$). Three different Co-related structures can be retrieved: (1) individual and well separated Co atoms, (2) Co clusters consisting of multiple atoms and (3) areas of Co/Ge intermixing layers. (b) STM image of 10 well separated individual Co atoms ($V_{\rm t} = +1.0 \, {\rm V}$, $I_{\rm t} = 100 \, {\rm pA}$). (c) Close-up view of an individual Co atom ($V_{\rm t} = +0.9 \, {\rm V}$, $I_{\rm t} = 300 \, {\rm pA}$). (d) Height profile taken along the line $AA$ across two different domains of the Ge surface that are separated by a Co/Ge intermixing layer.}
\label{Co_related_struct}
\end{figure}

%======  Classification of the new type structures on the Co/Ge(111) surface  ======%
The close-up view in Figure~\ref{Co_related_struct} comprises the three different types of Co related structures that can be observed after Co deposition on cold Ge(111)2$\times$1 surfaces: (i) well separated individual Co atoms [see label (1) in Fig.~\ref{Co_related_struct}~(a)]; (ii) Co clusters consisting out of multiple Co atoms [see label (2) in Fig.~\ref{Co_related_struct}~(a)]; (iii) Co/Ge ILs [see label (3) in Fig.~\ref{Co_related_struct}~(a)].

%=======================================================%
%======     Fig 3    " Co_related_struct "     =========%
%=======================================================%
%=======================================================%
%=======================================================%
%=======================================================%

%======  Single Co atoms  ======%
First, a significant fraction of the deposited Co atoms are retrieved as single, well separated atoms on the Ge(111)2$\times$1 surface. Around $13 \pm 6$\% of the deposited amount of Co atoms is observed as individual Co atoms, while near to $87 \pm 6$\% contributes to the formation of Co/Ge ILs and Co clusters. As illustrated in Fig.~\ref{Co_related_struct}~(b), single Co atoms appear as bright protrusions in the STM images and are located at the upper $\pi$-bonded chains at voltages above $ 0.7 \, {\rm V}$. High resolution STM images are presented in Figs.~\ref{Co_related_struct}~(c). Recently, we have shown that these Co atoms actually are not on top of the Ge(111)2$\times$1 surface, but they have penetrated into the Ge surface. They occupy quasi-stable positions inside the big 7-members Ge rings of the 2$\times$1 reconstruction, between the $3^{\rm rd}$ and the $4^{\rm th}$ atomic layer below the surface.~\cite{Muzychenko_Submitted_11} Embedding of deposited atoms into subsurface layers (even at RT) has already been demonstrated for Si~\cite{Uberuaga_PRL_00} and Ge~\cite{Lin_PRB_92,Gurlu_PRB_2004} surfaces. Here, the ``embedding" of Co atom in the Ge surface is found to significantly influence the local electronic structure, but it does not result in a modified surface reconstruction. As illustrated in Fig.~\ref{Co_related_struct}~(b) and (c), all individual Co atoms occupy identical positions. As a result, all the embedded Co atoms exhibit an identical electronic behavior and equally affect the electronic properties of the host $\pi$-bonded chain.~\cite{Muzychenko_Submitted_11}

%======  Co-clusters  ======%
Second, small Co clusters can be occasionally retrieved. As can be seen in Fig.~\ref{Co_related_struct}, the Co clusters have an arbitrary size and shape, from which it can be concluded that they consist of an arbitrary amount of Co atoms. The clusters exhibit strongly size dependent electronic properties that are different from those of individual Co atoms.

%=======================================================%
%======       Fig 4   " Co_IL_classes "        =========%
%=======================================================%
\begin{figure*}
\begin{center}\includegraphics{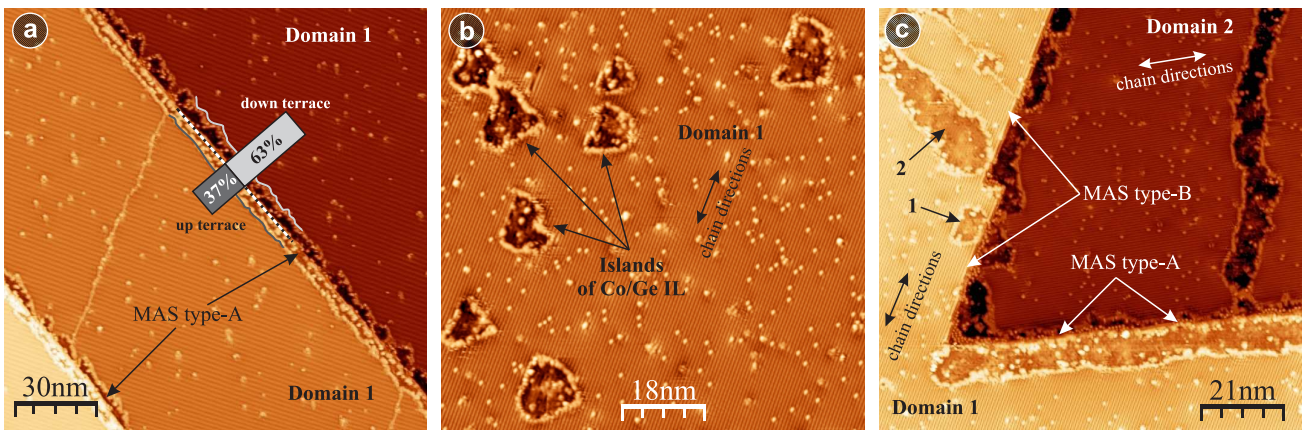}\end{center}
\caption{(Color online) Co/Ge ILs nucleate at three different surface locations, i.e., at (a) MASs, (b) on atomically flat Ge terraces at, e.g., vacancies or adatoms and (c) at DBs ($V_{\rm t} = +1.0 \, {\rm V}$, $I_{\rm t} = 15 \, {\rm pA}$).}
\label{Co_IL_classes}
\end{figure*}
%=======================================================%
%=======================================================%
%=======================================================%

%======  New Co/Ge intermixing layer   ======%
Third, larger areas of a new type of Co-related surface structures, the so-called Co/Ge ILs, can be observed after Co deposition on cold Ge surfaces [see the area enclosed by the two dotted black envelope curves, indicated by label (3), in Fig.~\ref{Co_related_struct}], which are calculated to comprise about $87 \pm 6 $\% of the amount of deposited Co atoms. From the height profile in Fig.~\ref{Co_related_struct}~(d), taken along the line $AA$, it can be seen that these Co/Ge ILs appear about $1.2 \, $\AA \ lower than the surrounding Ge terraces in STM images recorded at high positive voltages. This appearance differs from that of Ge adatom structures [see subsection~\ref{subsect:Ge(111)2x1}] and indicates a strongly modified electronic behavior of the Co/Ge IL when compared to the surrounding 2$\times$1 reconstruction. It is found that Co/Ge ILs nucleate at three different surface locations: (i) at MASs [see Figs.~\ref{Co_IL_classes}~(a) and (c)], (ii) at DBs [see Fig.~\ref{Co_related_struct} and Fig.~\ref{Co_IL_classes}~(c)] and (iii) on atomically flat Ge(111) terraces at, e.g., vacancies or adatoms [see Fig.~\ref{Co_IL_classes}~(b)]. It must be noted that, for type-A MASs, Co/Ge ILs are formed at both the upper and the lower terraces [Fig.~\ref{Co_IL_classes}~(a)], whereas for type-B MASs Co/Ge ILs are formed only at the lower terrace [see Fig.~\ref{Co_IL_classes}~(c)]. For DBs of both type-A and type-B Co/Ge ILs are formed at both sides of the DB. This can be clearly seen in Fig.~\ref{Co_related_struct}~(a), where a part of the unperturbed DB type-B can still be discerned at the bottom of STM image. In addition, Co/Ge ILs formed at a DB type-A between two surface domains with different directions of the 2$\times$1 reconstruction can also be observed in Fig.~\ref{Co_related_struct}~(a) [indicated by label (3)]. Finally, Co/Ge ILs can be formed on Ge(111)2$\times$1 terraces as well [Fig.~\ref{Co_IL_classes}~(b)]. Here, atomic size surface features such as Ga impurities or Ge vacancies can trap incoming diffusing Co atoms and act as nucleation points for the formation of two-dimensional Co/Ge ILs.

%--------------------------------------------------------------------------%
\subsection{Formation of the Co/Ge intermixing layer}
\label{subsect:Co/Ge_IL_formation}
%--------------------------------------------------------------------------%
Now we will discuss in more detail the formation mechanism of the Co/Ge ILs. Clearly, the size of the Co/Ge IL depends on the size of the neighboring single-domain Ge(111)2$\times$1 terrace, as can be observed in Fig.~\ref{Global_changes} and Fig.~\ref{Co_IL_classes}~(a). For example, for the case of Co/Ge ILs formed at the MASs in Fig.~\ref{Global_changes}, the width of the Co/Ge IL (measured normal to the MAS) at the narrow Ge terrace (2) is much smaller than the width of the Co/Ge IL formed at the wide Ge terrace (5). Note that all MASs in Fig.~\ref{Global_changes} are of type-A, except for the middle part of terrace (4). The (relative) total area of the Co/Ge IL on the upper and lower terraces at the MAS indicated in Fig.~\ref{Co_IL_classes}~(a) is around $37$\% and $63$\%, respectively. This corresponds well to the ratio of $1.9$ of the widths of the lower/upper terraces (which is determined from large scale STM images). Similarly, Ge terraces at MASs that have an even larger difference in terrace width were found to yield larger differences in the formed IL areas as well. A more detailed analysis of the dependence of Co/Ge ILs on the width of the neighboring (single-domain) Ge terrace indicates a nonlinear behavior. This can be explained by the fact that a fraction of the deposited Co atoms also contributes to Co/Ge ILs formed at DBs and on atomically flat Ge terraces at vacancies or adatoms [see Fig.~\ref{Global_changes} and Fig.~\ref{Co_IL_classes}~(b)].

Remarkably, as already noted above, Co/Ge ILs are not formed at the upper terraces of type-B MASs [see, e.g., the middle part of terrace (4) in Fig.~\ref{Global_changes}]. Considering the fact that the $\pi$-bonded chains of the 2$\times$1 reconstruction of the middle part of terrace (4) are oriented parallel to the neighboring MAS, this observation indicates that the Co/Ge ILs are formed by diffusion of deposited Co atoms along the $\pi$-bonded chains, until they are immobilized at, e.g. a MAS or a DB. The strong anisotropy of the Ge(111)2$\times$1 surface reconstruction with respect to the $[01\overline{1}]$ and $[2\overline{11}]$ directions~\cite{Northrup_PRB_83,Takeuchi_PRB_91} results in a preferred direction for the one-dimensional (1D) migration of the Co atoms, i.e., along the $\pi$-bonded chains. Indeed, it can be seen in Fig.~\ref{Co_IL_classes}~(c) that the upper terrace, forming both a type-A MAS and a type-B MAS, shows a continuous Co/Ge IL at the type-A MAS. The Co/Ge ILs at the type-B MAS [indicated by labels (1) and (2) in Fig.~\ref{Co_IL_classes}~(c)], however, are formed only locally [and may be caused by a vacancy/adatom (1) and a DB (2)] and most of the upper terrace remains unaltered near this type-B MAS. Ge adatoms that exist after cleavage of the Ge crystal are observed to show a similar diffusion behavior on (clean) Ge(111)2$\times$1 surfaces [see Fig.~\ref{Cleaved_Ge2x1_surface}~(c) and Section~\ref{subsect:Ge(111)2x1}].

%=======================================================%
%======     Fig 5     " Ge2x1+Co_warm_up "      ========%
%=======================================================%
\begin{figure*}
\begin{center}\includegraphics{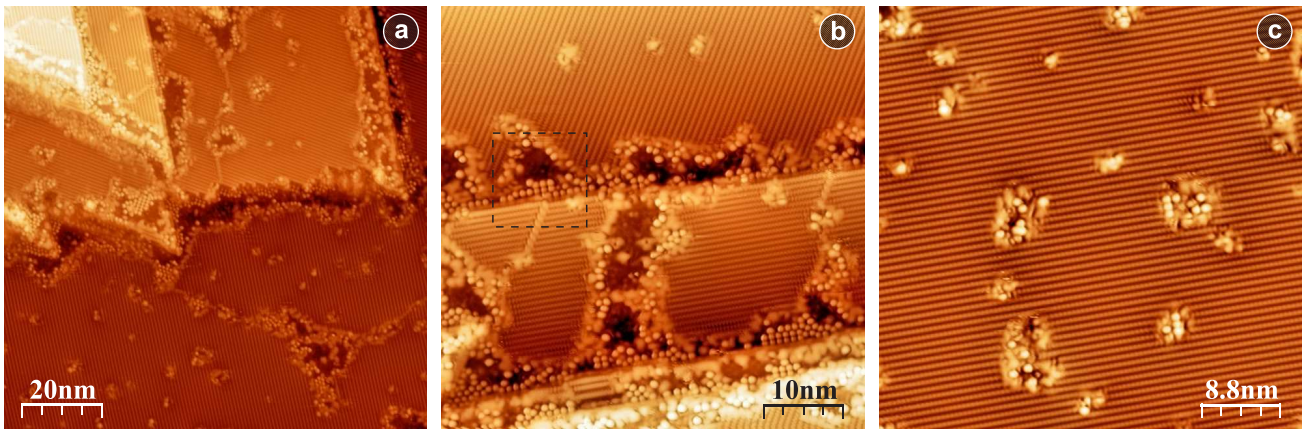}\end{center}
\caption{(Color online) Empty-states STM images of the Co/Ge(111)2$\times$1 system after warming the sample up to RT. Images are recorded at (a) $V_{\rm t} = +1.0 \, {\rm V}$ and $I_{\rm t} = 10 \, {\rm pA}$, (b) $V_{\rm t} = +0.87 \, {\rm V}$ and $I_{\rm t} = 35 \, {\rm pA}$ and (c) $V_{\rm t} = +1.0 \, {\rm V}$ and $I_{\rm t} = 20 \, {\rm pA}$.}
\label{Ge2x1+Co_warm_up}
\end{figure*}
%=======================================================%
%=======================================================%
%=======================================================%

Next, we investigated the influence of the substrate temperature on the diffusion behavior of the deposited Co atoms and hence on the formation of the Co/Ge ILs. For this purpose, the temperature of the Co/Ge(111)2$\times$1 sample (with $0.032$ MLs of Co) was increased up to RT for 24 hours, after which it was cooled down again to $4.5 \, {\rm K}$. Figures~\ref{Ge2x1+Co_warm_up}~(a)-(c) present three STM images of the Co/Ge(111)2$\times$1 surface recorded after this procedure. The individual Co atoms that were previously present at atomically flat Ge(111)2$\times$1 terraces [see Fig.~\ref{Co_related_struct} and Fig.~\ref{Co_IL_classes}] can no longer be observed. At the same time, the area of the Co/Ge ILs [Fig.~\ref{Ge2x1+Co_warm_up}~(a) and (b)] as well as the amount of Co clusters [Fig.~\ref{Ge2x1+Co_warm_up}~(c)] has increased. Since only $13$\% of the total amount of deposited Co atoms was retrieved as individual Co atoms on Ge terraces (before warming up the sample), the increase of the total Co/Ge IL area and the amount of Co clusters (after warming up the sample) cannot be accurately determined. Nevertheless, we can conclude that the mobility of the individual Co atoms was increased by increasing the sample temperature to RT. This is, however, remarkable since the individual Co atoms are actually embedded in the Ge surface after deposition, i.e., in between the $3^{\rm rd}$ and the $4^{\rm th}$ Ge layer.~\cite{Muzychenko_Submitted_11} During embedding of the Co atom in the Ge(111)2$\times$1 surface, a potential barrier of $\triangle E \simeq 0.5 \ {\rm eV}$ was overcome.~\cite{Muzychenko_Submitted_11} This implies that the Co atom cannot gain sufficient energy by warming the sample up to RT to return back to the surface ($E_{\rm RT} \simeq 25 \, {\rm meV}$). Migration of these embedded individual Co atoms at RT must therefore occur below/inside the Ge surface. The big 7-members Ge rings that exist along $\pi$-bonded chains in the $[01\overline{1}]$ direction seem to provide the only possible ``route" for this subsurface migration. As already discussed above, both Co/Ge ILs ($87$\%) and single Co atoms ($13$\%) are observed after Co atom deposition on cold Ge(111)2$\times$1 surfaces ($T_{\rm sample} \leq 80 \, {\rm K}$), from which it can be concluded that the majority of the deposited Co atoms already have a sufficiently high mobility at $\leq 80 \, {\rm K}$ and that they can diffuse along $\pi$-bonded chains in the $[01\overline{1}]$ direction. The sample temperature of $80 \, {\rm K}$ therefore appears to be a threshold value above which mostly Co/Ge ILs and Co clusters are formed, while the amount of individual Co atom increases with decreasing temperature below this threshold. Recently, it was shown that subsurface intermixing layers are formed at semiconductor surfaces by diffusion and migration of adsorbed atoms. For example, it was shown experimentally and theoretically that deposited Ge atoms already diffuse to the $4^{\rm th}$ subsurface Si layer of Si(100) surfaces at temperatures of $500 {\rm ^\circ C}$.~\cite{Uberuaga_PRL_00} Similarly, Si atoms deposited on Ge(100)2$\times$1 surfaces at RT were found to move below the Ge surface.~\cite{Lin_PRB_92} Also, previously it was experimentally shown the formation of intermixing layer due to Co deposition on Ge(111) surface at RT. Using MeV ion channeling and Auger electron spectrometry it was found that for the Co coverage up to 3 ML an intermixing layer are formed, possibly resulting a thin layer of CoGe,~\cite{Smith_JVSTA_89} which is in good agreement with our results.

In summary, we can state that the Co/Ge ILs are formed by subsurface accumulation of Co atoms at MASs, DBs and on atomically flat Ge terraces at, e.g., vacancies and adatoms due to 1D subsurface migration of Co atoms through the 7-members Ge rings of the $\pi$-bonded chains in between $3^{rd}$ and $4^{th}$ atomic layers, yielding spatially extended 2D Co/Ge ILs as observed in our STM experiments.

%--------------------------------------------------------------------------%
\subsection{Structure of the Co/Ge intermixing layer}
\label{subsect:Co/Ge_IL_struct}
%--------------------------------------------------------------------------%

\begin{figure}[b]
\begin{center}\includegraphics{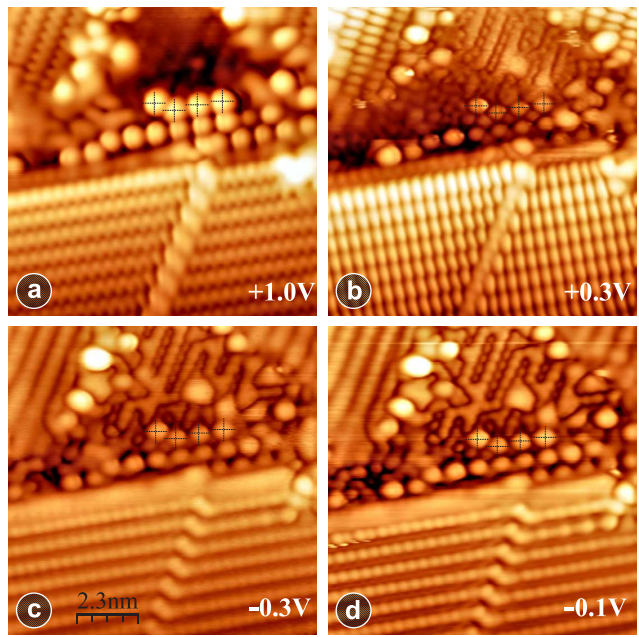}\end{center}
\caption{(Color online) (a, b) Empty and (c, d) filled states STM images of a Co/Ge IL formed at a type-B MAS, recorded at the location enclosed by the black dashed square in Fig.~\ref{Ge2x1+Co_warm_up}~(b). The tunneling voltage is indicated for each image [(a)-(c) $I_{\rm t} = 200 \, {\rm pA}$ and (d) $I_{\rm t} = 50 \, {\rm pA}$].} \label{Co/Ge_IL_bias_dep}
\end{figure}

At first sight, the Co/Ge ILs may seem to be rather disordered, without any clear periodicity. However, from analysis of voltage dependent STM images we found that Co/Ge ILs exhibit a highly ordered atomic structure that can be revealed only at low tunneling voltages. Figures~\ref{Co/Ge_IL_bias_dep}~(a,b) and (c,d) show empty and filled states STM images of a Co/Ge IL formed at a type-B MAS, recorded at the location enclosed by the black dashed square in Fig.~\ref{Ge2x1+Co_warm_up}~(b). As a guide for the eye, the same group of adatoms is indicated by black cross-hairs in Figs.~\ref{Co/Ge_IL_bias_dep}~(a)-(d).

%=======================================================%
%======     Fig 6    " Co/Ge_IL_bias_dep "      ========%
%=======================================================%
%=======================================================%
%=======================================================%
%=======================================================%

Remarkably, the Co/Ge IL and the surrounding 2$\times$1 reconstructed surface have about the same height at low voltages in the filled states regime close to $E_{\rm F}$ [Fig.~\ref{Co/Ge_IL_bias_dep}~(c) and (d)], while there appears to be a pronounced height difference of around $1.2$\AA\ at higher voltages [see height profile in Fig.~\ref{Co_IL_classes}~(d)]. This voltage dependent behavior can be related to a change of the electronic properties of the $\pi$-bonded chains at the surface, rather than to a change of the Ge(111) surface reconstruction. Moreover, the Co/Ge IL at the lower terrace in Fig.~\ref{Co/Ge_IL_bias_dep}~(d) (top part of the STM image) reveals, besides multiple adatoms, a locally highly ordered atomic structure with the same periodicity as that of the original 2$\times$1 reconstruction along the $\pi$-bonded chains. Along the $[2\overline{11}]$ direction, perpendicular to the $\pi$-bonded chains, it can be seen that this new Co/Ge IL reconstruction has a ``double periodicity".

%=======================================================%
%=========    Fig 7  " Co/Ge_IL_struct "    ============%
%=======================================================%
\begin{figure}
\begin{center}\includegraphics{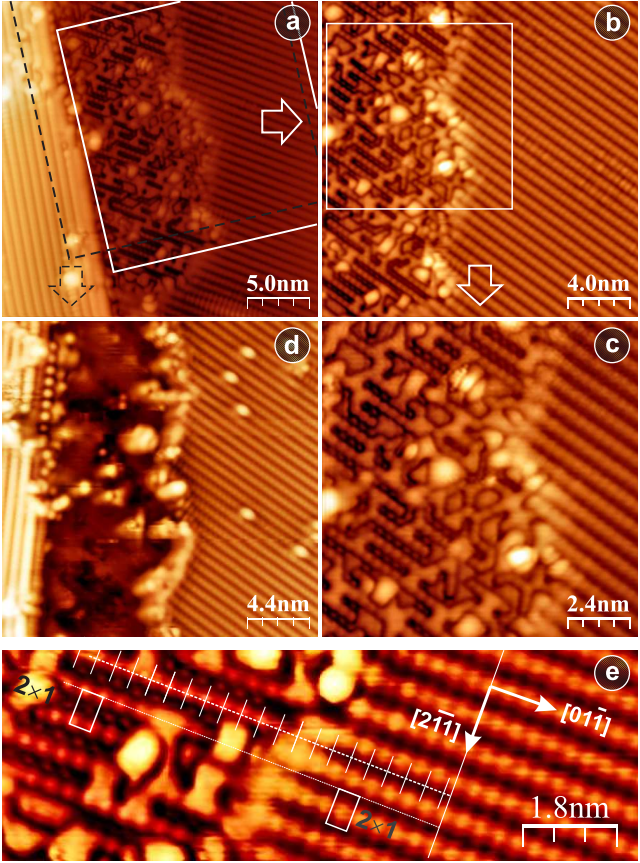}\end{center}
\caption{(Color online) (a) Filled states STM image of the Ge/Co IL formed at a type-B MAS ($V_{\rm t} = -0.4 \, {\rm V}$, $I_{\rm t} = 50 \, {\rm pA}$). (b), (c) and (d) Close-up views of the area enclosed by, respectively the white solid square in (a), the white solid square in (b) and the black dashed square in (a) ($V_{\rm t} = -0.4 \, {\rm V}$, $I_{\rm t} = 200 \, {\rm pA}$). (e) Atomic resolution STM image of the Co/Ge IL (left) and the Ge(111)2$\times$1 surface (right) at the type-B MAS ($V_{\rm t} = -0.4 \, {\rm V}$, $I_{\rm t} = 50 \, {\rm pA}$).}
\label{Co/Ge_IL_struct}
\end{figure}
%=======================================================%
%=======================================================%
%=======================================================%

\begin{figure}
\begin{center}\includegraphics{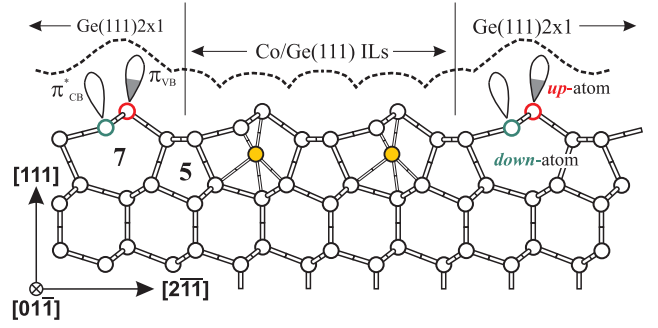}\end{center}
\caption{(Color online) Schematic side view of the 2$\times$1 reconstruction of the Ge(111) surface (left and right part) and Co/Ge(111) IL formed within two big 7-members Ge rings (middle part). The 7-members and 5-members Ge rings that are formed as a result of the surface reconstruction are indicated by the numbers 7 and 5, respectively. Two cobalt atoms are schematically shown in the center of the 7-members Ge rings as a solid yellow circles. }
\label{CoGe_schematic}
\end{figure}

Figure~\ref{Co/Ge_IL_struct}~(a) presents a filled states STM image of a Co/Ge IL formed at the lower terrace of another type-B MAS. This surface has not been warmed up to RT, so that single Co atoms can still be observed in the empty states STM image in Fig.~\ref{Co/Ge_IL_struct}~(d). A close-up view of the area confined by the white solid and black dashed squares in Fig.~\ref{Co/Ge_IL_struct}~(a) is shown in Fig.~\ref{Co/Ge_IL_struct}~(b) (filled state regime) and Fig.~\ref{Co/Ge_IL_struct}~(d) (empty state regime), respectively. Figures~\ref{Co/Ge_IL_struct}~(c) and (e) present two atomic resolution STM images of the Co/Ge IL and the surrounding Ge(111)2$\times$1 surface. Although the Co/Ge IL in general appears rather disordered, it locally also exhibits highly ordered atomic structures [Fig.~\ref{Co/Ge_IL_struct}~(c)], which indicates that these highly ordered structures of the Co/Ge ILs are already formed at LT, prior warming up to RT. Their formation is consistent with the migration mechanism discussed above, for which Co atoms migrate through the big 7-members Ge rings along $\pi$-bonded chains and accumulate at the MAS to form the Co/Ge IL. As a result, Co atoms are densely packed near the MAS inside the big 7-members Ge rings along the upper $\pi$-bonded chain (that is visualized in the STM images), while the 5-members Ge rings of the neighboring lower $\pi$-bonded chains remain empty. Figure~\ref{CoGe_schematic} show schematically atomic structure of the Ge(111)2$\times$1 surface as well as assumed structure of the Co/Ge IL formed by Co atoms embedded within big 7-members Ge rings. Therefore we can state that 2D Co/Ge ILs consist of alternating Co nanowires (with a periodicity of $6.9 \, \rm{nm}$) that are located in between the $3^{\rm rd}$ and $4^{\rm th}$ atomic layer underneath the Ge(111)2$\times$1 surface. Schematic structure [Fig.~\ref{CoGe_schematic}] does not include the relaxation of the Co/Ge IL (the model based on \emph{ab-initio} calculation will be published somewhere else) and assumes periodicity in the $[01\overline{1}]$ direction. In addition, the irregular structures on top of the Co/Ge ILs can be related to the Co atom deposition process. Already during the deposition process, deposited Co atoms are migrating and are hence forming the Co/Ge ILs. When a Co atom is then deposited on an already formed Co/Ge IL, it will reside here and is observed as a Co adatom on the Co/Ge IL in the STM images [see Fig.~\ref{Co/Ge_IL_bias_dep} and Fig.~\ref{Co/Ge_IL_struct}].

%========================================================%
%==========   Fig 8     " 2x1_schematic "    ============%
%========================================================%
%========================================================%
%========================================================%
%========================================================%

As discussed in Section~\ref{subsect:Ge(111)2x1}, the \emph{up}- and \emph{down}-atoms of the Ge(111)2$\times$1 reconstruction both have one dangling bond [see Fig.~\ref{CoGe_schematic}], from which the surface state bands are derived. The latter plays an important role in the visualization of the Ge surface in STM images.~\cite{Rohlfing_PRB_04} The presence of embedded Co atoms in the Ge surface implies Co-Ge bonding between the Co atom and Ge \emph{up}- and \emph{down}-atoms along $\pi$-bonded chains. As a result of the Co-Ge orbital hybridizations, both dangling bonds will be closed and the surface state will not be present any more at the upper $\pi$-bonded chains in the Co/Ge IL. Since, both of surface states play significant role in the STM image formation, then their ``absence" may result in a reduced (increased) contribution of the upper (lower) $\pi$-bonded chains to the tunneling current that constitutes the STM images at both polarities of the tunneling bias voltage. Schematically, the contour of constant local density of states integrated over an energy range from $E_{\rm F}$ to $E_{\rm F}+V_{\rm t}$ (where $E_{\rm F}$ is the Fermi level and $V_{\rm t}=\pm 0.5 \, {\rm eV}$ is the tunneling bias voltage) is shown in Fig.~\ref{CoGe_schematic} by black-dashed line. This implies that the upper (lower) $\pi$-bonded chains appear somewhat lower (higher) in the STM images when compared to the surrounding Ge surface, thereby causing the appearance of a double period in the STM images [Fig.~\ref{Co/Ge_IL_bias_dep}~(b)-(d) and Fig.~\ref{Co/Ge_IL_struct}~(a)-(c)] along the $[2\overline{11}]$ direction.

This can be seen more clearly in Fig.~\ref{Co/Ge_IL_struct}~(e). The higher atomic rows of the reconstructed Co/Ge IL in the $[01\overline{1}]$ direction follow the lower $\pi$-bonded chains of the Ge(111)2$\times$1 surface [see white dotted scale bar in Fig.~\ref{Co/Ge_IL_struct}~(e)]. The higher atomic rows of both regions are thus shifted by half of the total period along the $[2\overline{11}]$ direction with respect to each other. The unit surface cell (USC) of the Ge(111)2$\times$1 reconstruction is indicated in Fig.~\ref{Co/Ge_IL_struct}~(e), as well as the same USC shifted by $15$ periods along the $[01\overline{1}]$ direction. It can be seen that, after this shift, the local minima at the corners of the USC on the clean Ge(111)2$\times$1 surface now correspond to the local maxima of the Co/Ge IL. Similarly, the local maxima of the Ge(111)2$\times$1 reconstruction (i.e., the Ge \emph{up}-atoms) correspond to the local maxima of the (lower lying) ``secondary period" of the Co/Ge IL 2$\times$1 reconstruction (see scale bar in Fig.~\ref{Co/Ge_IL_struct}~(e)). Clearly, the new 2$\times$1 reconstruction of the Co/Ge IL matches perfectly the periodicity of both the upper and lower $\pi$-bonded chains of the clean Ge(111)2$\times$1 surface in the $[01\overline{1}]$ direction. However, as already discussed above, the periodicity doubles along the $[2\overline{11}]$ direction due to the modified contributions of the upper and lower $\pi$-bonded chains to the STM images. A full understanding of the Co induced reconstruction requires detailed theoretical calculations, which is beyond the scope of the current research and will be the point of future research.

%%%%%%%%%%%%%%%%%%%%%%%%%%%%%%%%%%%%%%%%%%%%%%%%%%%%%%%%%%%%%%%%%%%%%%%%%%%%
%%%%%%%%%%%%%%%%%%%%%%%%%%%%%%%%%%%%%%%%%%%%%%%%%%%%%%%%%%%%%%%%%%%%%%%%%%%%
%%%%%%%%%%%%%                   Conclusion                   %%%%%%%%%%%%%%%
%%%%%%%%%%%%%%%%%%%%%%%%%%%%%%%%%%%%%%%%%%%%%%%%%%%%%%%%%%%%%%%%%%%%%%%%%%%%
%%%%%%%%%%%%%%%%%%%%%%%%%%%%%%%%%%%%%%%%%%%%%%%%%%%%%%%%%%%%%%%%%%%%%%%%%%%%
\section{Conclusions}
By means of voltage dependent STM measurements at low temperature we have investigated the adsorption of Co atoms on cold cleaved Ge(111)2$\times$1 surfaces ($T \leq 80 \, {\rm K}$) in the submonolayer coverage range. Individual Co atoms, small Co clusters and Co/Ge ILs are retrieved on the Ge surface. We recently showed by DFT calculations that the individual Co atoms are embedded into the Ge(111)2$\times$1 surface, in between the $3^{\rm rd}$ and $4^{\rm th}$ atomic layers. Here we showed that the embedded Co atoms can migrate (even at low temperatures) underneath the Ge surface to MASs, DBs, vacancies and adatoms, which act as nucleation centers for Co/Ge ILs. If the substrate is warmed up to room temperature, the mobility of the individual Co atoms increases and they all contribute to the formation of the Co/Ge ILs. Here, these 2D Co/Ge ILs  are formed by accumulation of subsurface Co atoms that have migrated along the upper $\pi$-bonded chains. We have shown that these ILs (locally) have a highly ordered atomic structure, of which the periodicity coincides with that of the initial 2$\times$1 reconstruction in both $[01\overline{1}]$ and $[2\overline{11}]$ directions. The new reconstruction, however, has a double periodicity along the $[2\overline{11}]$ direction due to the modified electronic properties of the upper and lower $\pi$-bonded chains.

%\ack
\begin{acknowledgments}
The research in Moscow has been supported by the RFBR Grants. The research in Leuven has been supported by the Fund for Scientific Research - Flanders (FWO, Belgium) as well as by the Research Fund of the K.U. Leuven. K.S. is a postdoctoral researcher of the FWO. We thank S.V.~Savinov for his technical support and for providing the Ge crystals.
\end{acknowledgments}

%\section*{References}
%


\begin{thebibliography}{50}%
\makeatletter
\providecommand \@ifxundefined [1]{%
 \@ifx{#1\undefined}
}%
\providecommand \@ifnum [1]{%
 \ifnum #1\expandafter \@firstoftwo
 \else \expandafter \@secondoftwo
 \fi
}%
\providecommand \@ifx [1]{%
 \ifx #1\expandafter \@firstoftwo
 \else \expandafter \@secondoftwo
 \fi
}%
\providecommand \natexlab [1]{#1}%
\providecommand \enquote  [1]{``#1''}%
\providecommand \bibnamefont  [1]{#1}%
\providecommand \bibfnamefont [1]{#1}%
\providecommand \citenamefont [1]{#1}%
\providecommand \href@noop [0]{\@secondoftwo}%
\providecommand \href [0]{\begingroup \@sanitize@url \@href}%
\providecommand \@href[1]{\@@startlink{#1}\@@href}%
\providecommand \@@href[1]{\endgroup#1\@@endlink}%
\providecommand \@sanitize@url [0]{\catcode `\\12\catcode `\$12\catcode
  `\&12\catcode `\#12\catcode `\^12\catcode `\_12\catcode `\%12\relax}%
\providecommand \@@startlink[1]{}%
\providecommand \@@endlink[0]{}%
\providecommand \url  [0]{\begingroup\@sanitize@url \@url }%
\providecommand \@url [1]{\endgroup\@href {#1}{\urlprefix }}%
\providecommand \urlprefix  [0]{URL }%
\providecommand \Eprint [0]{\href }%
\@ifxundefined \urlstyle {%
  \providecommand \doi  [0]{\begingroup \@sanitize@url \@doi}%
  \providecommand \@doi [1]{\endgroup \@@startlink {\doibase
  #1}doi:\discretionary {}{}{}#1\@@endlink }%
}{%
  \providecommand \doi  [0]{doi:\discretionary{}{}{}\begingroup
  \urlstyle{rm}\Url }%
}%
\providecommand \doibase [0]{http://dx.doi.org/}%
\providecommand \Doi [0]{\begingroup \@sanitize@url \@Doi }%
\providecommand \@Doi  [1]{\endgroup\@@startlink{\doibase#1}\@@Doi}%
\providecommand \@@Doi [1]{#1\@@endlink}%
\providecommand \selectlanguage [0]{\@gobble}%
\providecommand \bibinfo  [0]{\@secondoftwo}%
\providecommand \bibfield  [0]{\@secondoftwo}%
\providecommand \translation [1]{[#1]}%
\providecommand \BibitemOpen [0]{}%
\providecommand \bibitemStop [0]{}%
\providecommand \bibitemNoStop [0]{.\EOS\space}%
\providecommand \EOS [0]{\spacefactor3000\relax}%
\providecommand \BibitemShut  [1]{\csname bibitem#1\endcsname}%
%</preamble>
\bibitem [{\citenamefont {Haider}\ \emph {et~al.}(2009)\citenamefont {Haider},
  \citenamefont {Pitters}, \citenamefont {DiLabio}, \citenamefont {Livadaru},
  \citenamefont {Mutus},\ and\ \citenamefont {Wolkow}}]{Haider_PRL_09}%
  \BibitemOpen
  \bibfield  {author} {\bibinfo {author} {\bibfnamefont {M.~B.}\ \bibnamefont
  {Haider}}, \bibinfo {author} {\bibfnamefont {J.~L.}\ \bibnamefont {Pitters}},
  \bibinfo {author} {\bibfnamefont {G.~A.}\ \bibnamefont {DiLabio}}, \bibinfo
  {author} {\bibfnamefont {L.}~\bibnamefont {Livadaru}}, \bibinfo {author}
  {\bibfnamefont {J.~Y.}\ \bibnamefont {Mutus}}, \ and\ \bibinfo {author}
  {\bibfnamefont {R.~A.}\ \bibnamefont {Wolkow}},\ }\Doi
  {10.1103/PhysRevLett.102.046805} {\bibfield  {journal} {\bibinfo  {journal}
  {Phys. Rev. Lett.},\ }\textbf {\bibinfo {volume} {102}},\ \bibinfo {pages}
  {046805} (\bibinfo {year} {2009})}\BibitemShut {NoStop}%
\bibitem [{\citenamefont {Tan}\ \emph {et~al.}(2009)\citenamefont {Tan},
  \citenamefont {Chan}, \citenamefont {Mttnen}, \citenamefont {Morello},
  \citenamefont {Yang}, \citenamefont {van Donkelaar}, \citenamefont {Alves},
  \citenamefont {Pirkkalainen}, \citenamefont {Jamieson}, \citenamefont
  {Clark},\ and\ \citenamefont {Dzurak}}]{Tan_NanoLett_09}%
  \BibitemOpen
  \bibfield  {author} {\bibinfo {author} {\bibfnamefont {K.~Y.}\ \bibnamefont
  {Tan}}, \bibinfo {author} {\bibfnamefont {K.~W.}\ \bibnamefont {Chan}},
  \bibinfo {author} {\bibfnamefont {M.}~\bibnamefont {Mttnen}}, \bibinfo
  {author} {\bibfnamefont {A.}~\bibnamefont {Morello}}, \bibinfo {author}
  {\bibfnamefont {C.}~\bibnamefont {Yang}}, \bibinfo {author} {\bibfnamefont
  {J.}~\bibnamefont {van Donkelaar}}, \bibinfo {author} {\bibfnamefont
  {A.}~\bibnamefont {Alves}}, \bibinfo {author} {\bibfnamefont {J.-M.}\
  \bibnamefont {Pirkkalainen}}, \bibinfo {author} {\bibfnamefont {D.~N.}\
  \bibnamefont {Jamieson}}, \bibinfo {author} {\bibfnamefont {R.~G.}\
  \bibnamefont {Clark}}, \ and\ \bibinfo {author} {\bibfnamefont {A.~S.}\
  \bibnamefont {Dzurak}},\ }\Doi {10.1021/nl901635j} {\bibfield  {journal}
  {\bibinfo  {journal} {Nano Lett.},\ }\textbf {\bibinfo {volume} {10}},\
  \bibinfo {pages} {11} (\bibinfo {year} {2009})}\BibitemShut {NoStop}%
\bibitem [{\citenamefont {Parks}\ \emph {et~al.}(2010)\citenamefont {Parks},
  \citenamefont {Champagne}, \citenamefont {Costi}, \citenamefont {Shum},
  \citenamefont {Pasupathy}, \citenamefont {Neuscamman}, \citenamefont
  {Flores-Torres}, \citenamefont {Cornaglia}, \citenamefont {Aligia},
  \citenamefont {Balseiro}, \citenamefont {Chan}, \citenamefont {Abruna},\ and\
  \citenamefont {Ralph}}]{Parks_Science_10}%
  \BibitemOpen
  \bibfield  {author} {\bibinfo {author} {\bibfnamefont {J.~J.}\ \bibnamefont
  {Parks}}, \bibinfo {author} {\bibfnamefont {A.~R.}\ \bibnamefont
  {Champagne}}, \bibinfo {author} {\bibfnamefont {T.~A.}\ \bibnamefont
  {Costi}}, \bibinfo {author} {\bibfnamefont {W.~W.}\ \bibnamefont {Shum}},
  \bibinfo {author} {\bibfnamefont {A.~N.}\ \bibnamefont {Pasupathy}}, \bibinfo
  {author} {\bibfnamefont {E.}~\bibnamefont {Neuscamman}}, \bibinfo {author}
  {\bibfnamefont {S.}~\bibnamefont {Flores-Torres}}, \bibinfo {author}
  {\bibfnamefont {P.~S.}\ \bibnamefont {Cornaglia}}, \bibinfo {author}
  {\bibfnamefont {A.~A.}\ \bibnamefont {Aligia}}, \bibinfo {author}
  {\bibfnamefont {C.~A.}\ \bibnamefont {Balseiro}}, \bibinfo {author}
  {\bibfnamefont {G.~K.-L.}\ \bibnamefont {Chan}}, \bibinfo {author}
  {\bibfnamefont {H.~D.}\ \bibnamefont {Abruna}}, \ and\ \bibinfo {author}
  {\bibfnamefont {D.~C.}\ \bibnamefont {Ralph}},\ }\Doi
  {10.1126/science.1186874} {\bibfield  {journal} {\bibinfo  {journal}
  {Science},\ }\textbf {\bibinfo {volume} {328}},\ \bibinfo {pages} {1370}
  (\bibinfo {year} {2010})}\BibitemShut {NoStop}%
\bibitem [{\citenamefont {Chui}\ \emph {et~al.}(2003)\citenamefont {Chui},
  \citenamefont {Gopalakrishnan}, \citenamefont {Griffin}, \citenamefont
  {Plummer},\ and\ \citenamefont {Saraswat}}]{Chui_APL_03}%
  \BibitemOpen
  \bibfield  {author} {\bibinfo {author} {\bibfnamefont {C.~O.}\ \bibnamefont
  {Chui}}, \bibinfo {author} {\bibfnamefont {K.}~\bibnamefont
  {Gopalakrishnan}}, \bibinfo {author} {\bibfnamefont {P.~B.}\ \bibnamefont
  {Griffin}}, \bibinfo {author} {\bibfnamefont {J.~D.}\ \bibnamefont
  {Plummer}}, \ and\ \bibinfo {author} {\bibfnamefont {K.~C.}\ \bibnamefont
  {Saraswat}},\ }\Doi {doi:10.1063/1.1618382} {\bibfield  {journal} {\bibinfo
  {journal} {Appl. Phys. Lett.},\ }\textbf {\bibinfo {volume} {83}},\ \bibinfo
  {pages} {3275} (\bibinfo {year} {2003})}\BibitemShut {NoStop}%
\bibitem [{\citenamefont {Bracht}\ \emph {et~al.}(2009)\citenamefont {Bracht},
  \citenamefont {Schneider}, \citenamefont {Klug}, \citenamefont {Liao},
  \citenamefont {Hansen}, \citenamefont {Haller}, \citenamefont {Larsen},
  \citenamefont {Bougeard}, \citenamefont {Posselt},\ and\ \citenamefont
  {W\"undisch}}]{Bracht_PRL_09}%
  \BibitemOpen
  \bibfield  {author} {\bibinfo {author} {\bibfnamefont {H.}~\bibnamefont
  {Bracht}}, \bibinfo {author} {\bibfnamefont {S.}~\bibnamefont {Schneider}},
  \bibinfo {author} {\bibfnamefont {J.~N.}\ \bibnamefont {Klug}}, \bibinfo
  {author} {\bibfnamefont {C.~Y.}\ \bibnamefont {Liao}}, \bibinfo {author}
  {\bibfnamefont {J.~L.}\ \bibnamefont {Hansen}}, \bibinfo {author}
  {\bibfnamefont {E.~E.}\ \bibnamefont {Haller}}, \bibinfo {author}
  {\bibfnamefont {A.~N.}\ \bibnamefont {Larsen}}, \bibinfo {author}
  {\bibfnamefont {D.}~\bibnamefont {Bougeard}}, \bibinfo {author}
  {\bibfnamefont {M.}~\bibnamefont {Posselt}}, \ and\ \bibinfo {author}
  {\bibfnamefont {C.}~\bibnamefont {W\"undisch}},\ }\Doi
  {10.1103/PhysRevLett.103.255501} {\bibfield  {journal} {\bibinfo  {journal}
  {Phys. Rev. Lett.},\ }\textbf {\bibinfo {volume} {103}},\ \bibinfo {pages}
  {255501} (\bibinfo {year} {2009})}\BibitemShut {NoStop}%
\bibitem [{\citenamefont {Wundisch}\ \emph {et~al.}(2009)\citenamefont
  {Wundisch}, \citenamefont {Posselt}, \citenamefont {Schmidt}, \citenamefont
  {Heera}, \citenamefont {Schumann}, \citenamefont {Mucklich}, \citenamefont
  {Grotzschel}, \citenamefont {Skorupa}, \citenamefont {Clarysse},
  \citenamefont {Simoen},\ and\ \citenamefont {Hortenbach}}]{Wundisch_APL_09}%
  \BibitemOpen
  \bibfield  {author} {\bibinfo {author} {\bibfnamefont {C.}~\bibnamefont
  {Wundisch}}, \bibinfo {author} {\bibfnamefont {M.}~\bibnamefont {Posselt}},
  \bibinfo {author} {\bibfnamefont {B.}~\bibnamefont {Schmidt}}, \bibinfo
  {author} {\bibfnamefont {V.}~\bibnamefont {Heera}}, \bibinfo {author}
  {\bibfnamefont {T.}~\bibnamefont {Schumann}}, \bibinfo {author}
  {\bibfnamefont {A.}~\bibnamefont {Mucklich}}, \bibinfo {author}
  {\bibfnamefont {R.}~\bibnamefont {Grotzschel}}, \bibinfo {author}
  {\bibfnamefont {W.}~\bibnamefont {Skorupa}}, \bibinfo {author} {\bibfnamefont
  {T.}~\bibnamefont {Clarysse}}, \bibinfo {author} {\bibfnamefont
  {E.}~\bibnamefont {Simoen}}, \ and\ \bibinfo {author} {\bibfnamefont
  {H.}~\bibnamefont {Hortenbach}},\ }\Doi {10.1063/1.3276770} {\bibfield
  {journal} {\bibinfo  {journal} {Appl. Phys. Lett.},\ }\textbf {\bibinfo
  {volume} {95}},\ \bibinfo {eid} {252107} (\bibinfo {year}
  {2009})}\BibitemShut {NoStop}%
\bibitem [{\citenamefont {Sze}\ and\ \citenamefont
  {Irvin}(1968)}]{Sze_SolStEl_68}%
  \BibitemOpen
  \bibfield  {author} {\bibinfo {author} {\bibfnamefont {S.~M.}\ \bibnamefont
  {Sze}}\ and\ \bibinfo {author} {\bibfnamefont {J.~C.}\ \bibnamefont
  {Irvin}},\ }\Doi {10.1016/0038-1101(68)90012-9} {\bibfield  {journal}
  {\bibinfo  {journal} {Sol. St. Elect.},\ }\textbf {\bibinfo {volume} {11}},\
  \bibinfo {pages} {599} (\bibinfo {year} {1968})}\BibitemShut {NoStop}%
\bibitem [{\citenamefont {Rowe}\ \emph {et~al.}(1975)\citenamefont {Rowe},
  \citenamefont {Christman},\ and\ \citenamefont {Margaritondo}}]{Rowe_PRL_75}%
  \BibitemOpen
  \bibfield  {author} {\bibinfo {author} {\bibfnamefont {J.~E.}\ \bibnamefont
  {Rowe}}, \bibinfo {author} {\bibfnamefont {S.~B.}\ \bibnamefont {Christman}},
  \ and\ \bibinfo {author} {\bibfnamefont {G.}~\bibnamefont {Margaritondo}},\
  }\Doi {10.1103/PhysRevLett.35.1471} {\bibfield  {journal} {\bibinfo
  {journal} {Phys. Rev. Lett.},\ }\textbf {\bibinfo {volume} {35}},\ \bibinfo
  {pages} {1471} (\bibinfo {year} {1975})}\BibitemShut {NoStop}%
\bibitem [{\citenamefont {Chang}\ \emph {et~al.}(1989)\citenamefont {Chang},
  \citenamefont {Hwu}, \citenamefont {Hansen}, \citenamefont {Zanini},\ and\
  \citenamefont {Margaritondo}}]{Chang_PRL_89}%
  \BibitemOpen
  \bibfield  {author} {\bibinfo {author} {\bibfnamefont {Y.}~\bibnamefont
  {Chang}}, \bibinfo {author} {\bibfnamefont {Y.}~\bibnamefont {Hwu}}, \bibinfo
  {author} {\bibfnamefont {J.}~\bibnamefont {Hansen}}, \bibinfo {author}
  {\bibfnamefont {F.}~\bibnamefont {Zanini}}, \ and\ \bibinfo {author}
  {\bibfnamefont {G.}~\bibnamefont {Margaritondo}},\ }\Doi
  {10.1103/PhysRevLett.63.1845} {\bibfield  {journal} {\bibinfo  {journal}
  {Phys. Rev. Lett.},\ }\textbf {\bibinfo {volume} {63}},\ \bibinfo {pages}
  {1845} (\bibinfo {year} {1989})}\BibitemShut {NoStop}%
\bibitem [{\citenamefont {Lee}\ \emph {et~al.}(2006)\citenamefont {Lee},
  \citenamefont {Oh}, \citenamefont {Tseng}, \citenamefont {Jammy},\ and\
  \citenamefont {Huff}}]{Lee_MatTod_06}%
  \BibitemOpen
  \bibfield  {author} {\bibinfo {author} {\bibfnamefont {B.~H.}\ \bibnamefont
  {Lee}}, \bibinfo {author} {\bibfnamefont {J.}~\bibnamefont {Oh}}, \bibinfo
  {author} {\bibfnamefont {H.~H.}\ \bibnamefont {Tseng}}, \bibinfo {author}
  {\bibfnamefont {R.}~\bibnamefont {Jammy}}, \ and\ \bibinfo {author}
  {\bibfnamefont {H.}~\bibnamefont {Huff}},\ }\Doi {DOI:
  10.1016/S1369-7021(06)71541-3} {\bibfield  {journal} {\bibinfo  {journal}
  {Mat. Today},\ }\textbf {\bibinfo {volume} {9}},\ \bibinfo {pages} {32 }
  (\bibinfo {year} {2006})}\BibitemShut {NoStop}%
\bibitem [{\citenamefont {Tse}\ and\ \citenamefont
  {Robertson}(2007)}]{Koon-Yiu_PRL_07}%
  \BibitemOpen
  \bibfield  {author} {\bibinfo {author} {\bibfnamefont {K.-Y.}\ \bibnamefont
  {Tse}}\ and\ \bibinfo {author} {\bibfnamefont {J.}~\bibnamefont
  {Robertson}},\ }\Doi {10.1103/PhysRevLett.99.086805} {\bibfield  {journal}
  {\bibinfo  {journal} {Phys. Rev. Lett.},\ }\textbf {\bibinfo {volume} {99}},\
  \bibinfo {pages} {086805} (\bibinfo {year} {2007})}\BibitemShut {NoStop}%
\bibitem [{\citenamefont {Zhu}\ \emph {et~al.}(2005)\citenamefont {Zhu},
  \citenamefont {Li}, \citenamefont {Lee}, \citenamefont {Li}, \citenamefont
  {Du}, \citenamefont {Singh}, \citenamefont {Zhu}, \citenamefont {Chin},\ and\
  \citenamefont {Kwong}}]{Zhu_IEEE_05}%
  \BibitemOpen
  \bibfield  {author} {\bibinfo {author} {\bibfnamefont {S.}~\bibnamefont
  {Zhu}}, \bibinfo {author} {\bibfnamefont {R.}~\bibnamefont {Li}}, \bibinfo
  {author} {\bibfnamefont {S.~J.}\ \bibnamefont {Lee}}, \bibinfo {author}
  {\bibfnamefont {M.~F.}\ \bibnamefont {Li}}, \bibinfo {author} {\bibfnamefont
  {A.}~\bibnamefont {Du}}, \bibinfo {author} {\bibfnamefont {J.}~\bibnamefont
  {Singh}}, \bibinfo {author} {\bibfnamefont {C.}~\bibnamefont {Zhu}}, \bibinfo
  {author} {\bibfnamefont {A.}~\bibnamefont {Chin}}, \ and\ \bibinfo {author}
  {\bibfnamefont {D.~L.}\ \bibnamefont {Kwong}},\ }\Doi
  {10.1109/LED.2004.841462} {\bibfield  {journal} {\bibinfo  {journal} {Elect.
  Dev. Lett., IEEE},\ }\textbf {\bibinfo {volume} {26}},\ \bibinfo {pages} {81}
  (\bibinfo {year} {2005})}\BibitemShut {NoStop}%
\bibitem [{\citenamefont {Park}\ \emph {et~al.}(2007)\citenamefont {Park},
  \citenamefont {Lee}, \citenamefont {Lee}, \citenamefont {Ko}, \citenamefont
  {Kwak}, \citenamefont {Yang},\ and\ \citenamefont {Kim}}]{Park_JES_07}%
  \BibitemOpen
  \bibfield  {author} {\bibinfo {author} {\bibfnamefont {K.}~\bibnamefont
  {Park}}, \bibinfo {author} {\bibfnamefont {B.~H.}\ \bibnamefont {Lee}},
  \bibinfo {author} {\bibfnamefont {D.}~\bibnamefont {Lee}}, \bibinfo {author}
  {\bibfnamefont {D.-H.}\ \bibnamefont {Ko}}, \bibinfo {author} {\bibfnamefont
  {K.~H.}\ \bibnamefont {Kwak}}, \bibinfo {author} {\bibfnamefont {C.-W.}\
  \bibnamefont {Yang}}, \ and\ \bibinfo {author} {\bibfnamefont
  {H.}~\bibnamefont {Kim}},\ }\Doi {10.1149/1.2732164} {\bibfield  {journal}
  {\bibinfo  {journal} {J. Electrochem. Soc.},\ }\textbf {\bibinfo {volume}
  {154}},\ \bibinfo {pages} {H557} (\bibinfo {year} {2007})}\BibitemShut
  {NoStop}%
\bibitem [{\citenamefont {Brunco}\ \emph {et~al.}(2008)\citenamefont {Brunco},
  \citenamefont {Jaeger}, \citenamefont {Eneman}, \citenamefont {Mitard},
  \citenamefont {Hellings}, \citenamefont {Satta}, \citenamefont {Terzieva},
  \citenamefont {Souriau}, \citenamefont {Leys}, \citenamefont {Pourtois},
  \citenamefont {Houssa}, \citenamefont {Winderickx}, \citenamefont {Vrancken},
  \citenamefont {Sioncke}, \citenamefont {Opsomer}, \citenamefont {Nicholas},
  \citenamefont {Caymax}, \citenamefont {Stesmans}, \citenamefont
  {Steenbergen}, \citenamefont {Mertens}, \citenamefont {Meuris},\ and\
  \citenamefont {Heyns}}]{Brunco_JES_08}%
  \BibitemOpen
  \bibfield  {author} {\bibinfo {author} {\bibfnamefont {D.~P.}\ \bibnamefont
  {Brunco}}, \bibinfo {author} {\bibfnamefont {B.~D.}\ \bibnamefont {Jaeger}},
  \bibinfo {author} {\bibfnamefont {G.}~\bibnamefont {Eneman}}, \bibinfo
  {author} {\bibfnamefont {J.}~\bibnamefont {Mitard}}, \bibinfo {author}
  {\bibfnamefont {G.}~\bibnamefont {Hellings}}, \bibinfo {author}
  {\bibfnamefont {A.}~\bibnamefont {Satta}}, \bibinfo {author} {\bibfnamefont
  {V.}~\bibnamefont {Terzieva}}, \bibinfo {author} {\bibfnamefont
  {L.}~\bibnamefont {Souriau}}, \bibinfo {author} {\bibfnamefont {F.~E.}\
  \bibnamefont {Leys}}, \bibinfo {author} {\bibfnamefont {G.}~\bibnamefont
  {Pourtois}}, \bibinfo {author} {\bibfnamefont {M.}~\bibnamefont {Houssa}},
  \bibinfo {author} {\bibfnamefont {G.}~\bibnamefont {Winderickx}}, \bibinfo
  {author} {\bibfnamefont {E.}~\bibnamefont {Vrancken}}, \bibinfo {author}
  {\bibfnamefont {S.}~\bibnamefont {Sioncke}}, \bibinfo {author} {\bibfnamefont
  {K.}~\bibnamefont {Opsomer}}, \bibinfo {author} {\bibfnamefont
  {G.}~\bibnamefont {Nicholas}}, \bibinfo {author} {\bibfnamefont
  {M.}~\bibnamefont {Caymax}}, \bibinfo {author} {\bibfnamefont
  {A.}~\bibnamefont {Stesmans}}, \bibinfo {author} {\bibfnamefont {J.~V.}\
  \bibnamefont {Steenbergen}}, \bibinfo {author} {\bibfnamefont {P.~W.}\
  \bibnamefont {Mertens}}, \bibinfo {author} {\bibfnamefont {M.}~\bibnamefont
  {Meuris}}, \ and\ \bibinfo {author} {\bibfnamefont {M.~M.}\ \bibnamefont
  {Heyns}},\ }\Doi {10.1149/1.2919115} {\bibfield  {journal} {\bibinfo
  {journal} {J. Electrochem. Soc.},\ }\textbf {\bibinfo {volume} {155}},\
  \bibinfo {pages} {H552} (\bibinfo {year} {2008})}\BibitemShut {NoStop}%
\bibitem [{\citenamefont {Gaudet}\ \emph
  {et~al.}(2006){\natexlab{a}}\citenamefont {Gaudet}, \citenamefont
  {Detavernier}, \citenamefont {Kellock}, \citenamefont {Desjardins},\ and\
  \citenamefont {Lavoie}}]{Gaudet_JVSTA_06}%
  \BibitemOpen
  \bibfield  {author} {\bibinfo {author} {\bibfnamefont {S.}~\bibnamefont
  {Gaudet}}, \bibinfo {author} {\bibfnamefont {C.}~\bibnamefont {Detavernier}},
  \bibinfo {author} {\bibfnamefont {A.~J.}\ \bibnamefont {Kellock}}, \bibinfo
  {author} {\bibfnamefont {P.}~\bibnamefont {Desjardins}}, \ and\ \bibinfo
  {author} {\bibfnamefont {C.}~\bibnamefont {Lavoie}},\ }\Doi
  {10.1116/1.2191861} {\bibfield  {journal} {\bibinfo  {journal} {J. Vac. Sci.
  Technol. A},\ }\textbf {\bibinfo {volume} {24}},\ \bibinfo {pages} {474}
  (\bibinfo {year} {2006}{\natexlab{a}})}\BibitemShut {NoStop}%
\bibitem [{\citenamefont {Ashburn}\ \emph {et~al.}(1993)\citenamefont
  {Ashburn}, \citenamefont {\"{O}zt\"{u}rk}, \citenamefont {Harris},\ and\
  \citenamefont {Maher}}]{Ashburn_JAP_93}%
  \BibitemOpen
  \bibfield  {author} {\bibinfo {author} {\bibfnamefont {S.~P.}\ \bibnamefont
  {Ashburn}}, \bibinfo {author} {\bibfnamefont {M.~C.}\ \bibnamefont
  {\"{O}zt\"{u}rk}}, \bibinfo {author} {\bibfnamefont {G.}~\bibnamefont
  {Harris}}, \ and\ \bibinfo {author} {\bibfnamefont {D.~M.}\ \bibnamefont
  {Maher}},\ }\Doi {10.1063/1.354387} {\bibfield  {journal} {\bibinfo
  {journal} {J. Appl. Phys.},\ }\textbf {\bibinfo {volume} {74}},\ \bibinfo
  {pages} {4455} (\bibinfo {year} {1993})}\BibitemShut {NoStop}%
\bibitem [{\citenamefont {Gaudet}\ \emph
  {et~al.}(2006){\natexlab{b}}\citenamefont {Gaudet}, \citenamefont
  {Detavernier}, \citenamefont {Lavoie},\ and\ \citenamefont
  {Desjardins}}]{Gaudet_JAP_06}%
  \BibitemOpen
  \bibfield  {author} {\bibinfo {author} {\bibfnamefont {S.}~\bibnamefont
  {Gaudet}}, \bibinfo {author} {\bibfnamefont {C.}~\bibnamefont {Detavernier}},
  \bibinfo {author} {\bibfnamefont {C.}~\bibnamefont {Lavoie}}, \ and\ \bibinfo
  {author} {\bibfnamefont {P.}~\bibnamefont {Desjardins}},\ }\Doi
  {10.1063/1.2219080} {\bibfield  {journal} {\bibinfo  {journal} {J. Appl.
  Phys.},\ }\textbf {\bibinfo {volume} {100}},\ \bibinfo {eid} {034306}
  (\bibinfo {year} {2006}{\natexlab{b}})}\BibitemShut {NoStop}%
\bibitem [{\citenamefont {Opsomer}\ \emph {et~al.}(2007)\citenamefont
  {Opsomer}, \citenamefont {Deduytsche}, \citenamefont {Detavernier},
  \citenamefont {Meirhaeghe}, \citenamefont {Lauwers}, \citenamefont {Maex},\
  and\ \citenamefont {Lavoie}}]{Opsomer_APL_07}%
  \BibitemOpen
  \bibfield  {author} {\bibinfo {author} {\bibfnamefont {K.}~\bibnamefont
  {Opsomer}}, \bibinfo {author} {\bibfnamefont {D.}~\bibnamefont {Deduytsche}},
  \bibinfo {author} {\bibfnamefont {C.}~\bibnamefont {Detavernier}}, \bibinfo
  {author} {\bibfnamefont {R.~L.~V.}\ \bibnamefont {Meirhaeghe}}, \bibinfo
  {author} {\bibfnamefont {A.}~\bibnamefont {Lauwers}}, \bibinfo {author}
  {\bibfnamefont {K.}~\bibnamefont {Maex}}, \ and\ \bibinfo {author}
  {\bibfnamefont {C.}~\bibnamefont {Lavoie}},\ }\Doi {10.1063/1.2431781}
  {\bibfield  {journal} {\bibinfo  {journal} {Appl. Phys. Lett.},\ }\textbf
  {\bibinfo {volume} {90}},\ \bibinfo {eid} {031906} (\bibinfo {year}
  {2007})}\BibitemShut {NoStop}%
\bibitem [{\citenamefont {Chi}\ \emph {et~al.}(2005)\citenamefont {Chi},
  \citenamefont {Lee}, \citenamefont {Chua}, \citenamefont {Lee}, \citenamefont
  {Ashok},\ and\ \citenamefont {Kwong}}]{Chi_JAP_05}%
  \BibitemOpen
  \bibfield  {author} {\bibinfo {author} {\bibfnamefont {D.~Z.}\ \bibnamefont
  {Chi}}, \bibinfo {author} {\bibfnamefont {R.~T.~P.}\ \bibnamefont {Lee}},
  \bibinfo {author} {\bibfnamefont {S.~J.}\ \bibnamefont {Chua}}, \bibinfo
  {author} {\bibfnamefont {S.~J.}\ \bibnamefont {Lee}}, \bibinfo {author}
  {\bibfnamefont {S.}~\bibnamefont {Ashok}}, \ and\ \bibinfo {author}
  {\bibfnamefont {D.-L.}\ \bibnamefont {Kwong}},\ }\Doi {10.1063/1.1923162}
  {\bibfield  {journal} {\bibinfo  {journal} {J. Appl. Phys.},\ }\textbf
  {\bibinfo {volume} {97}},\ \bibinfo {eid} {113706} (\bibinfo {year}
  {2005})}\BibitemShut {NoStop}%
\bibitem [{\citenamefont {Simoen}\ \emph {et~al.}(2006)\citenamefont {Simoen},
  \citenamefont {Opsomer}, \citenamefont {Claeys}, \citenamefont {Maex},
  \citenamefont {Detavernier}, \citenamefont {Meirhaeghe}, \citenamefont
  {Forment},\ and\ \citenamefont {Clauws}}]{Simoen_APL_06}%
  \BibitemOpen
  \bibfield  {author} {\bibinfo {author} {\bibfnamefont {E.}~\bibnamefont
  {Simoen}}, \bibinfo {author} {\bibfnamefont {K.}~\bibnamefont {Opsomer}},
  \bibinfo {author} {\bibfnamefont {C.}~\bibnamefont {Claeys}}, \bibinfo
  {author} {\bibfnamefont {K.}~\bibnamefont {Maex}}, \bibinfo {author}
  {\bibfnamefont {C.}~\bibnamefont {Detavernier}}, \bibinfo {author}
  {\bibfnamefont {R.~L.~V.}\ \bibnamefont {Meirhaeghe}}, \bibinfo {author}
  {\bibfnamefont {S.}~\bibnamefont {Forment}}, \ and\ \bibinfo {author}
  {\bibfnamefont {P.}~\bibnamefont {Clauws}},\ }\Doi {10.1063/1.2199615}
  {\bibfield  {journal} {\bibinfo  {journal} {Appl. Phys. Lett.},\ }\textbf
  {\bibinfo {volume} {88}},\ \bibinfo {eid} {183506} (\bibinfo {year}
  {2006})}\BibitemShut {NoStop}%
\bibitem [{\citenamefont {Simoen}\ \emph {et~al.}(2008)\citenamefont {Simoen},
  \citenamefont {Opsomer}, \citenamefont {Claeys}, \citenamefont {Maex},
  \citenamefont {Detavernier}, \citenamefont {Meirhaeghe},\ and\ \citenamefont
  {Clauws}}]{Simoen_JAP_08}%
  \BibitemOpen
  \bibfield  {author} {\bibinfo {author} {\bibfnamefont {E.}~\bibnamefont
  {Simoen}}, \bibinfo {author} {\bibfnamefont {K.}~\bibnamefont {Opsomer}},
  \bibinfo {author} {\bibfnamefont {C.}~\bibnamefont {Claeys}}, \bibinfo
  {author} {\bibfnamefont {K.}~\bibnamefont {Maex}}, \bibinfo {author}
  {\bibfnamefont {C.}~\bibnamefont {Detavernier}}, \bibinfo {author}
  {\bibfnamefont {R.~L.~V.}\ \bibnamefont {Meirhaeghe}}, \ and\ \bibinfo
  {author} {\bibfnamefont {P.}~\bibnamefont {Clauws}},\ }\Doi
  {10.1063/1.2956708} {\bibfield  {journal} {\bibinfo  {journal} {J. Appl.
  Phys.},\ }\textbf {\bibinfo {volume} {104}},\ \bibinfo {eid} {023705}
  (\bibinfo {year} {2008})}\BibitemShut {NoStop}%
\bibitem [{\citenamefont {Kittl}\ \emph {et~al.}(2008)\citenamefont {Kittl},
  \citenamefont {Opsomer}, \citenamefont {Torregiani}, \citenamefont
  {Demeurisse}, \citenamefont {Mertens}, \citenamefont {Brunco}, \citenamefont
  {Dal},\ and\ \citenamefont {Lauwers}}]{Kittl_MSEB_08}%
  \BibitemOpen
  \bibfield  {author} {\bibinfo {author} {\bibfnamefont {J.}~\bibnamefont
  {Kittl}}, \bibinfo {author} {\bibfnamefont {K.}~\bibnamefont {Opsomer}},
  \bibinfo {author} {\bibfnamefont {C.}~\bibnamefont {Torregiani}}, \bibinfo
  {author} {\bibfnamefont {C.}~\bibnamefont {Demeurisse}}, \bibinfo {author}
  {\bibfnamefont {S.}~\bibnamefont {Mertens}}, \bibinfo {author} {\bibfnamefont
  {D.}~\bibnamefont {Brunco}}, \bibinfo {author} {\bibfnamefont {M.~V.}\
  \bibnamefont {Dal}}, \ and\ \bibinfo {author} {\bibfnamefont
  {A.}~\bibnamefont {Lauwers}},\ }\Doi {DOI: 10.1016/j.mseb.2008.09.033}
  {\bibfield  {journal} {\bibinfo  {journal} {Mat. Sci. Engin. B},\ }\textbf
  {\bibinfo {volume} {154-155}},\ \bibinfo {pages} {144 } (\bibinfo {year}
  {2008})}\BibitemShut {NoStop}%
\bibitem [{\citenamefont {Wolf}\ \emph {et~al.}(2001)\citenamefont {Wolf},
  \citenamefont {Awschalom}, \citenamefont {Buhrman}, \citenamefont {Daughton},
  \citenamefont {von Molnár}, \citenamefont {Roukes}, \citenamefont
  {Chtchelkanova},\ and\ \citenamefont {Treger}}]{Wolf_Sci_01}%
  \BibitemOpen
  \bibfield  {author} {\bibinfo {author} {\bibfnamefont {S.~A.}\ \bibnamefont
  {Wolf}}, \bibinfo {author} {\bibfnamefont {D.~D.}\ \bibnamefont {Awschalom}},
  \bibinfo {author} {\bibfnamefont {R.~A.}\ \bibnamefont {Buhrman}}, \bibinfo
  {author} {\bibfnamefont {J.~M.}\ \bibnamefont {Daughton}}, \bibinfo {author}
  {\bibfnamefont {S.}~\bibnamefont {von Molnár}}, \bibinfo {author}
  {\bibfnamefont {M.~L.}\ \bibnamefont {Roukes}}, \bibinfo {author}
  {\bibfnamefont {A.~Y.}\ \bibnamefont {Chtchelkanova}}, \ and\ \bibinfo
  {author} {\bibfnamefont {D.~M.}\ \bibnamefont {Treger}},\ }\Doi
  {10.1126/science.1065389} {\bibfield  {journal} {\bibinfo  {journal}
  {Science},\ }\textbf {\bibinfo {volume} {294}},\ \bibinfo {pages} {1488}
  (\bibinfo {year} {2001})}\BibitemShut {NoStop}%
\bibitem [{\citenamefont {\ifmmode \check{Z}\else
  \v{Z}\fi{}uti\ifmmode~\acute{c}\else \'{c}\fi{}}\ \emph
  {et~al.}(2004)\citenamefont {\ifmmode \check{Z}\else
  \v{Z}\fi{}uti\ifmmode~\acute{c}\else \'{c}\fi{}}, \citenamefont {Fabian},\
  and\ \citenamefont {Das~Sarma}}]{Zutic_RMP_04}%
  \BibitemOpen
  \bibfield  {author} {\bibinfo {author} {\bibfnamefont {I.}~\bibnamefont
  {\ifmmode \check{Z}\else \v{Z}\fi{}uti\ifmmode~\acute{c}\else \'{c}\fi{}}},
  \bibinfo {author} {\bibfnamefont {J.}~\bibnamefont {Fabian}}, \ and\ \bibinfo
  {author} {\bibfnamefont {S.}~\bibnamefont {Das~Sarma}},\ }\Doi
  {10.1103/RevModPhys.76.323} {\bibfield  {journal} {\bibinfo  {journal} {Rev.
  Mod. Phys.},\ }\textbf {\bibinfo {volume} {76}},\ \bibinfo {pages} {323}
  (\bibinfo {year} {2004})}\BibitemShut {NoStop}%
\bibitem [{\citenamefont {Ryan}\ \emph {et~al.}(2004)\citenamefont {Ryan},
  \citenamefont {Winarski}, \citenamefont {Keavney}, \citenamefont {Freeland},
  \citenamefont {Rosenberg}, \citenamefont {Park},\ and\ \citenamefont
  {Falco}}]{Ryan_PRB_04}%
  \BibitemOpen
  \bibfield  {author} {\bibinfo {author} {\bibfnamefont {P.}~\bibnamefont
  {Ryan}}, \bibinfo {author} {\bibfnamefont {R.~P.}\ \bibnamefont {Winarski}},
  \bibinfo {author} {\bibfnamefont {D.~J.}\ \bibnamefont {Keavney}}, \bibinfo
  {author} {\bibfnamefont {J.~W.}\ \bibnamefont {Freeland}}, \bibinfo {author}
  {\bibfnamefont {R.~A.}\ \bibnamefont {Rosenberg}}, \bibinfo {author}
  {\bibfnamefont {S.}~\bibnamefont {Park}}, \ and\ \bibinfo {author}
  {\bibfnamefont {C.~M.}\ \bibnamefont {Falco}},\ }\Doi
  {10.1103/PhysRevB.69.054416} {\bibfield  {journal} {\bibinfo  {journal}
  {Phys. Rev. B},\ }\textbf {\bibinfo {volume} {69}},\ \bibinfo {pages}
  {054416} (\bibinfo {year} {2004})}\BibitemShut {NoStop}%
\bibitem [{\citenamefont {Mello}\ \emph {et~al.}(1997)\citenamefont {Mello},
  \citenamefont {Murarka}, \citenamefont {Lu},\ and\ \citenamefont
  {Lee}}]{Mello_JAP_97}%
  \BibitemOpen
  \bibfield  {author} {\bibinfo {author} {\bibfnamefont {K.~E.}\ \bibnamefont
  {Mello}}, \bibinfo {author} {\bibfnamefont {S.~P.}\ \bibnamefont {Murarka}},
  \bibinfo {author} {\bibfnamefont {T.-M.}\ \bibnamefont {Lu}}, \ and\ \bibinfo
  {author} {\bibfnamefont {S.~L.}\ \bibnamefont {Lee}},\ }\Doi
  {10.1063/1.365323} {\bibfield  {journal} {\bibinfo  {journal} {J. Appl.
  Phys.},\ }\textbf {\bibinfo {volume} {81}},\ \bibinfo {pages} {7261}
  (\bibinfo {year} {1997})}\BibitemShut {NoStop}%
\bibitem [{\citenamefont {Dhar}\ and\ \citenamefont
  {Kulkarni}(1998)}]{Dhar_TSF_98}%
  \BibitemOpen
  \bibfield  {author} {\bibinfo {author} {\bibfnamefont {S.}~\bibnamefont
  {Dhar}}\ and\ \bibinfo {author} {\bibfnamefont {V.~N.}\ \bibnamefont
  {Kulkarni}},\ }\Doi {DOI: 10.1016/S0040-6090(98)00607-5} {\bibfield
  {journal} {\bibinfo  {journal} {Thin Solid Films},\ }\textbf {\bibinfo
  {volume} {333}},\ \bibinfo {pages} {20 } (\bibinfo {year}
  {1998})}\BibitemShut {NoStop}%
\bibitem [{\citenamefont {Goldfarb}\ and\ \citenamefont
  {Briggs}(2000)}]{Goldfarb_JMR_00}%
  \BibitemOpen
  \bibfield  {author} {\bibinfo {author} {\bibfnamefont {I.}~\bibnamefont
  {Goldfarb}}\ and\ \bibinfo {author} {\bibfnamefont {G.~A.~D.}\ \bibnamefont
  {Briggs}},\ }\Doi {10.1557/JMR.2001.0103} {\bibfield  {journal} {\bibinfo
  {journal} {J. Mater. Res},\ }\textbf {\bibinfo {volume} {16}},\ \bibinfo
  {pages} {744} (\bibinfo {year} {2000})}\BibitemShut {NoStop}%
\bibitem [{\citenamefont {Tsay}\ \emph {et~al.}(2004)\citenamefont {Tsay},
  \citenamefont {Nieh}, \citenamefont {Yao}, \citenamefont {Chen},\ and\
  \citenamefont {Cheng}}]{Tsay_SS_04}%
  \BibitemOpen
  \bibfield  {author} {\bibinfo {author} {\bibfnamefont {J.~S.}\ \bibnamefont
  {Tsay}}, \bibinfo {author} {\bibfnamefont {H.~Y.}\ \bibnamefont {Nieh}},
  \bibinfo {author} {\bibfnamefont {Y.~D.}\ \bibnamefont {Yao}}, \bibinfo
  {author} {\bibfnamefont {Y.~T.}\ \bibnamefont {Chen}}, \ and\ \bibinfo
  {author} {\bibfnamefont {W.~C.}\ \bibnamefont {Cheng}},\ }\Doi {DOI:
  10.1016/j.susc.2004.06.123} {\bibfield  {journal} {\bibinfo  {journal} {Surf.
  Sci.},\ }\textbf {\bibinfo {volume} {566-568}},\ \bibinfo {pages} {226 }
  (\bibinfo {year} {2004})}\BibitemShut {NoStop}%
\bibitem [{\citenamefont {Sun}\ \emph {et~al.}(2005)\citenamefont {Sun},
  \citenamefont {Chen}, \citenamefont {Pan}, \citenamefont {Chi}, \citenamefont
  {Nath},\ and\ \citenamefont {Foo}}]{Sun_APL_05}%
  \BibitemOpen
  \bibfield  {author} {\bibinfo {author} {\bibfnamefont {H.~P.}\ \bibnamefont
  {Sun}}, \bibinfo {author} {\bibfnamefont {Y.~B.}\ \bibnamefont {Chen}},
  \bibinfo {author} {\bibfnamefont {X.~Q.}\ \bibnamefont {Pan}}, \bibinfo
  {author} {\bibfnamefont {D.~Z.}\ \bibnamefont {Chi}}, \bibinfo {author}
  {\bibfnamefont {R.}~\bibnamefont {Nath}}, \ and\ \bibinfo {author}
  {\bibfnamefont {Y.~L.}\ \bibnamefont {Foo}},\ }\Doi {10.1063/1.2135387}
  {\bibfield  {journal} {\bibinfo  {journal} {Appl. Phys. Lett.},\ }\textbf
  {\bibinfo {volume} {87}},\ \bibinfo {eid} {211909} (\bibinfo {year}
  {2005})}\BibitemShut {NoStop}%
\bibitem [{\citenamefont {Chang}\ \emph {et~al.}(2006)\citenamefont {Chang},
  \citenamefont {Tsay}, \citenamefont {Chiou}, \citenamefont {Huang},
  \citenamefont {Chan},\ and\ \citenamefont {Yao}}]{Chang_JAP_06}%
  \BibitemOpen
  \bibfield  {author} {\bibinfo {author} {\bibfnamefont {H.~W.}\ \bibnamefont
  {Chang}}, \bibinfo {author} {\bibfnamefont {J.~S.}\ \bibnamefont {Tsay}},
  \bibinfo {author} {\bibfnamefont {Y.~L.}\ \bibnamefont {Chiou}}, \bibinfo
  {author} {\bibfnamefont {K.~T.}\ \bibnamefont {Huang}}, \bibinfo {author}
  {\bibfnamefont {W.~Y.}\ \bibnamefont {Chan}}, \ and\ \bibinfo {author}
  {\bibfnamefont {Y.~D.}\ \bibnamefont {Yao}},\ }\Doi {10.1063/1.2176315}
  {\bibfield  {journal} {\bibinfo  {journal} {J. Appl. Phys.},\ }\textbf
  {\bibinfo {volume} {99}},\ \bibinfo {eid} {08J705} (\bibinfo {year}
  {2006})}\BibitemShut {NoStop}%
\bibitem [{\citenamefont {{K. Sell}}\ \emph {et~al.}(2007)\citenamefont {{K.
  Sell}}, \citenamefont {{A. Kleibert}}, \citenamefont {{V. v. Oeynhausen}},\
  and\ \citenamefont {{K.-H. Meiwes-Broer}}}]{Sell_EPJD_07}%
  \BibitemOpen
  \bibfield  {author} {\bibinfo {author} {\bibnamefont {{K. Sell}}}, \bibinfo
  {author} {\bibnamefont {{A. Kleibert}}}, \bibinfo {author} {\bibnamefont {{V.
  v. Oeynhausen}}}, \ and\ \bibinfo {author} {\bibnamefont {{K.-H.
  Meiwes-Broer}}},\ }\Doi {10.1140/epjd/e2007-00213-7} {\bibfield  {journal}
  {\bibinfo  {journal} {Eur. Phys. J. D},\ }\textbf {\bibinfo {volume} {45}},\
  \bibinfo {pages} {433} (\bibinfo {year} {2007})}\BibitemShut {NoStop}%
\bibitem [{\citenamefont {Park}\ \emph {et~al.}(2009)\citenamefont {Park},
  \citenamefont {An}, \citenamefont {Lee}, \citenamefont {Yang}, \citenamefont
  {Lee},\ and\ \citenamefont {Kim}}]{Park_JECS_09}%
  \BibitemOpen
  \bibfield  {author} {\bibinfo {author} {\bibfnamefont {K.}~\bibnamefont
  {Park}}, \bibinfo {author} {\bibfnamefont {C.-H.}\ \bibnamefont {An}},
  \bibinfo {author} {\bibfnamefont {M.~S.}\ \bibnamefont {Lee}}, \bibinfo
  {author} {\bibfnamefont {C.-W.}\ \bibnamefont {Yang}}, \bibinfo {author}
  {\bibfnamefont {H.-J.}\ \bibnamefont {Lee}}, \ and\ \bibinfo {author}
  {\bibfnamefont {H.}~\bibnamefont {Kim}},\ }\Doi {10.1149/1.3071634}
  {\bibfield  {journal} {\bibinfo  {journal} {J. Electrochem. Soc.},\ }\textbf
  {\bibinfo {volume} {156}},\ \bibinfo {pages} {H229} (\bibinfo {year}
  {2009})}\BibitemShut {NoStop}%
\bibitem [{\citenamefont {Keyser}\ \emph {et~al.}(2010)\citenamefont {Keyser},
  \citenamefont {Meirhaeghe}, \citenamefont {Detavernier}, \citenamefont
  {Jordan-Sweet},\ and\ \citenamefont {Lavoie}}]{Keyser_JECS_10}%
  \BibitemOpen
  \bibfield  {author} {\bibinfo {author} {\bibfnamefont {K.~D.}\ \bibnamefont
  {Keyser}}, \bibinfo {author} {\bibfnamefont {R.~L.~V.}\ \bibnamefont
  {Meirhaeghe}}, \bibinfo {author} {\bibfnamefont {C.}~\bibnamefont
  {Detavernier}}, \bibinfo {author} {\bibfnamefont {J.}~\bibnamefont
  {Jordan-Sweet}}, \ and\ \bibinfo {author} {\bibfnamefont {C.}~\bibnamefont
  {Lavoie}},\ }\Doi {10.1149/1.3294702} {\bibfield  {journal} {\bibinfo
  {journal} {J. Electrochem. Soc.},\ }\textbf {\bibinfo {volume} {157}},\
  \bibinfo {pages} {H395} (\bibinfo {year} {2010})}\BibitemShut {NoStop}%
\bibitem [{\citenamefont {Pandey}(1981)}]{Pandey_PRL_81}%
  \BibitemOpen
  \bibfield  {author} {\bibinfo {author} {\bibfnamefont {K.~C.}\ \bibnamefont
  {Pandey}},\ }\Doi {10.1103/PhysRevLett.47.1913} {\bibfield  {journal}
  {\bibinfo  {journal} {Phys. Rev. Lett.},\ }\textbf {\bibinfo {volume} {47}},\
  \bibinfo {pages} {1913} (\bibinfo {year} {1981})}\BibitemShut {NoStop}%
\bibitem [{\citenamefont {Northrup}\ and\ \citenamefont
  {Cohen}(1983)}]{Northrup_PRB_83}%
  \BibitemOpen
  \bibfield  {author} {\bibinfo {author} {\bibfnamefont {J.~E.}\ \bibnamefont
  {Northrup}}\ and\ \bibinfo {author} {\bibfnamefont {M.~L.}\ \bibnamefont
  {Cohen}},\ }\Doi {10.1103/PhysRevB.27.6553} {\bibfield  {journal} {\bibinfo
  {journal} {Phys. Rev. B},\ }\textbf {\bibinfo {volume} {27}},\ \bibinfo
  {pages} {6553} (\bibinfo {year} {1983})}\BibitemShut {NoStop}%
\bibitem [{\citenamefont {Barth}\ \emph {et~al.}(1990)\citenamefont {Barth},
  \citenamefont {Brune}, \citenamefont {Ertl},\ and\ \citenamefont
  {Behm}}]{Barth_JV_90}%
  \BibitemOpen
  \bibfield  {author} {\bibinfo {author} {\bibfnamefont {J.~V.}\ \bibnamefont
  {Barth}}, \bibinfo {author} {\bibfnamefont {H.}~\bibnamefont {Brune}},
  \bibinfo {author} {\bibfnamefont {G.}~\bibnamefont {Ertl}}, \ and\ \bibinfo
  {author} {\bibfnamefont {R.~J.}\ \bibnamefont {Behm}},\ }\Doi
  {10.1103/PhysRevB.42.9307} {\bibfield  {journal} {\bibinfo  {journal} {Phys.
  Rev. B},\ }\textbf {\bibinfo {volume} {42}},\ \bibinfo {pages} {9307}
  (\bibinfo {year} {1990})}\BibitemShut {NoStop}%
\bibitem [{\citenamefont {Schouteden}\ \emph {et~al.}(2009)\citenamefont
  {Schouteden}, \citenamefont {Lijnen}, \citenamefont {Muzychenko},
  \citenamefont {Ceulemans}, \citenamefont {Chibotaru}, \citenamefont
  {Lievens},\ and\ \citenamefont {Haesendonck}}]{Schouteden_Nanotech_09}%
  \BibitemOpen
  \bibfield  {author} {\bibinfo {author} {\bibfnamefont {K.}~\bibnamefont
  {Schouteden}}, \bibinfo {author} {\bibfnamefont {E.}~\bibnamefont {Lijnen}},
  \bibinfo {author} {\bibfnamefont {D.~A.}\ \bibnamefont {Muzychenko}},
  \bibinfo {author} {\bibfnamefont {A.}~\bibnamefont {Ceulemans}}, \bibinfo
  {author} {\bibfnamefont {L.~F.}\ \bibnamefont {Chibotaru}}, \bibinfo {author}
  {\bibfnamefont {P.}~\bibnamefont {Lievens}}, \ and\ \bibinfo {author}
  {\bibfnamefont {C.~V.}\ \bibnamefont {Haesendonck}},\ }\href@noop {}
  {\bibfield  {journal} {\bibinfo  {journal} {Nanotech.},\ }\textbf {\bibinfo
  {volume} {20}},\ \bibinfo {pages} {395401} (\bibinfo {year}
  {2009})}\BibitemShut {NoStop}%
\bibitem [{\citenamefont {Horcas}\ \emph {et~al.}(2007)\citenamefont {Horcas},
  \citenamefont {Fernandez}, \citenamefont {Gomez-Rodriguez}, \citenamefont
  {Colchero}, \citenamefont {Gomez-Herrero},\ and\ \citenamefont
  {Baro}}]{WXsM}%
  \BibitemOpen
  \bibfield  {author} {\bibinfo {author} {\bibfnamefont {I.}~\bibnamefont
  {Horcas}}, \bibinfo {author} {\bibfnamefont {R.}~\bibnamefont {Fernandez}},
  \bibinfo {author} {\bibfnamefont {J.~M.}\ \bibnamefont {Gomez-Rodriguez}},
  \bibinfo {author} {\bibfnamefont {J.}~\bibnamefont {Colchero}}, \bibinfo
  {author} {\bibfnamefont {J.}~\bibnamefont {Gomez-Herrero}}, \ and\ \bibinfo
  {author} {\bibfnamefont {A.~M.}\ \bibnamefont {Baro}},\ }\Doi
  {10.1063/1.2432410} {\bibfield  {journal} {\bibinfo  {journal} {Rev. Sci.
  Inst.},\ }\textbf {\bibinfo {volume} {78}},\ \bibinfo {eid} {013705}
  (\bibinfo {year} {2007})}\BibitemShut {NoStop}%
\bibitem [{\citenamefont {Muzychenko}\ \emph {et~al.}(2010)\citenamefont
  {Muzychenko}, \citenamefont {Savinov}, \citenamefont {Mantsevich},
  \citenamefont {Maslova}, \citenamefont {Panov}, \citenamefont {Schouteden},\
  and\ \citenamefont {Van~Haesendonck}}]{Muzychenko_PRB_10}%
  \BibitemOpen
  \bibfield  {author} {\bibinfo {author} {\bibfnamefont {D.~A.}\ \bibnamefont
  {Muzychenko}}, \bibinfo {author} {\bibfnamefont {S.~V.}\ \bibnamefont
  {Savinov}}, \bibinfo {author} {\bibfnamefont {V.~N.}\ \bibnamefont
  {Mantsevich}}, \bibinfo {author} {\bibfnamefont {N.~S.}\ \bibnamefont
  {Maslova}}, \bibinfo {author} {\bibfnamefont {V.~I.}\ \bibnamefont {Panov}},
  \bibinfo {author} {\bibfnamefont {K.}~\bibnamefont {Schouteden}}, \ and\
  \bibinfo {author} {\bibfnamefont {C.}~\bibnamefont {Van~Haesendonck}},\ }\Doi
  {10.1103/PhysRevB.81.035313} {\bibfield  {journal} {\bibinfo  {journal}
  {Phys. Rev. B},\ }\textbf {\bibinfo {volume} {81}},\ \bibinfo {pages}
  {035313} (\bibinfo {year} {2010})}\BibitemShut {NoStop}%
\bibitem [{\citenamefont {Feenstra}\ and\ \citenamefont
  {Slavin}(1991)}]{Feenstra_SS_91}%
  \BibitemOpen
  \bibfield  {author} {\bibinfo {author} {\bibfnamefont {R.}~\bibnamefont
  {Feenstra}}\ and\ \bibinfo {author} {\bibfnamefont {A.}~\bibnamefont
  {Slavin}},\ }\Doi {DOI: 10.1016/0039-6028(91)91023-Q} {\bibfield  {journal}
  {\bibinfo  {journal} {Surf. Sci.},\ }\textbf {\bibinfo {volume} {251-252}},\
  \bibinfo {pages} {401 } (\bibinfo {year} {1991})}\BibitemShut {NoStop}%
\bibitem [{\citenamefont {Feenstra}\ \emph {et~al.}(2001)\citenamefont
  {Feenstra}, \citenamefont {Meyer}, \citenamefont {Moresco},\ and\
  \citenamefont {Rieder}}]{Feenstra_PRB_01}%
  \BibitemOpen
  \bibfield  {author} {\bibinfo {author} {\bibfnamefont {R.~M.}\ \bibnamefont
  {Feenstra}}, \bibinfo {author} {\bibfnamefont {G.}~\bibnamefont {Meyer}},
  \bibinfo {author} {\bibfnamefont {F.}~\bibnamefont {Moresco}}, \ and\
  \bibinfo {author} {\bibfnamefont {K.~H.}\ \bibnamefont {Rieder}},\ }\Doi
  {10.1103/PhysRevB.64.081306} {\bibfield  {journal} {\bibinfo  {journal}
  {Phys. Rev. B},\ }\textbf {\bibinfo {volume} {64}},\ \bibinfo {pages}
  {081306} (\bibinfo {year} {2001})}\BibitemShut {NoStop}%
\bibitem [{\citenamefont {Takeuchi}\ \emph {et~al.}(1991)\citenamefont
  {Takeuchi}, \citenamefont {Selloni}, \citenamefont {Shkrebtii},\ and\
  \citenamefont {Tosatti}}]{Takeuchi_PRB_91}%
  \BibitemOpen
  \bibfield  {author} {\bibinfo {author} {\bibfnamefont {N.}~\bibnamefont
  {Takeuchi}}, \bibinfo {author} {\bibfnamefont {A.}~\bibnamefont {Selloni}},
  \bibinfo {author} {\bibfnamefont {A.~I.}\ \bibnamefont {Shkrebtii}}, \ and\
  \bibinfo {author} {\bibfnamefont {E.}~\bibnamefont {Tosatti}},\ }\Doi
  {10.1103/PhysRevB.44.13611} {\bibfield  {journal} {\bibinfo  {journal} {Phys.
  Rev. B},\ }\textbf {\bibinfo {volume} {44}},\ \bibinfo {pages} {13611}
  (\bibinfo {year} {1991})}\BibitemShut {NoStop}%
\bibitem [{\citenamefont {Muzychenko}\ \emph {et~al.}(2011)\citenamefont
  {Muzychenko}, \citenamefont {Schouteden}, \citenamefont {Houssa},
  \citenamefont {Savinov},\ and\ \citenamefont
  {Van~Haesendonck}}]{Muzychenko_Submitted_11}%
  \BibitemOpen
  \bibfield  {author} {\bibinfo {author} {\bibfnamefont {D.~A.}\ \bibnamefont
  {Muzychenko}}, \bibinfo {author} {\bibfnamefont {K.}~\bibnamefont
  {Schouteden}}, \bibinfo {author} {\bibfnamefont {M.}~\bibnamefont {Houssa}},
  \bibinfo {author} {\bibfnamefont {S.~V.}\ \bibnamefont {Savinov}}, \ and\
  \bibinfo {author} {\bibfnamefont {C.}~\bibnamefont {Van~Haesendonck}},\
  }\href@noop {} {\bibfield  {journal} {\bibinfo  {journal} {Submitted to Phys.
  Rev. B (available online: http://arxiv.org/abs/****.****)}} (\bibinfo {year}
  {2011})}\BibitemShut {NoStop}%
\bibitem [{\citenamefont {Einaga}\ \emph {et~al.}(1998)\citenamefont {Einaga},
  \citenamefont {Hirayama},\ and\ \citenamefont {Takayanagi}}]{Einaga_PRB_98}%
  \BibitemOpen
  \bibfield  {author} {\bibinfo {author} {\bibfnamefont {Y.}~\bibnamefont
  {Einaga}}, \bibinfo {author} {\bibfnamefont {H.}~\bibnamefont {Hirayama}}, \
  and\ \bibinfo {author} {\bibfnamefont {K.}~\bibnamefont {Takayanagi}},\ }\Doi
  {10.1103/PhysRevB.57.15567} {\bibfield  {journal} {\bibinfo  {journal} {Phys.
  Rev. B},\ }\textbf {\bibinfo {volume} {57}},\ \bibinfo {pages} {15567}
  (\bibinfo {year} {1998})}\BibitemShut {NoStop}%
\bibitem [{\citenamefont {Uberuaga}\ \emph {et~al.}(2000)\citenamefont
  {Uberuaga}, \citenamefont {Leskovar}, \citenamefont {Smith}, \citenamefont
  {J\'onsson},\ and\ \citenamefont {Olmstead}}]{Uberuaga_PRL_00}%
  \BibitemOpen
  \bibfield  {author} {\bibinfo {author} {\bibfnamefont {B.~P.}\ \bibnamefont
  {Uberuaga}}, \bibinfo {author} {\bibfnamefont {M.}~\bibnamefont {Leskovar}},
  \bibinfo {author} {\bibfnamefont {A.~P.}\ \bibnamefont {Smith}}, \bibinfo
  {author} {\bibfnamefont {H.}~\bibnamefont {J\'onsson}}, \ and\ \bibinfo
  {author} {\bibfnamefont {M.}~\bibnamefont {Olmstead}},\ }\Doi
  {10.1103/PhysRevLett.84.2441} {\bibfield  {journal} {\bibinfo  {journal}
  {Phys. Rev. Lett.},\ }\textbf {\bibinfo {volume} {84}},\ \bibinfo {pages}
  {2441} (\bibinfo {year} {2000})}\BibitemShut {NoStop}%
\bibitem [{\citenamefont {Lin}\ \emph {et~al.}(1992)\citenamefont {Lin},
  \citenamefont {Miller},\ and\ \citenamefont {Chiang}}]{Lin_PRB_92}%
  \BibitemOpen
  \bibfield  {author} {\bibinfo {author} {\bibfnamefont {D.-S.}\ \bibnamefont
  {Lin}}, \bibinfo {author} {\bibfnamefont {T.}~\bibnamefont {Miller}}, \ and\
  \bibinfo {author} {\bibfnamefont {T.-C.}\ \bibnamefont {Chiang}},\ }\Doi
  {10.1103/PhysRevB.45.11415} {\bibfield  {journal} {\bibinfo  {journal} {Phys.
  Rev. B},\ }\textbf {\bibinfo {volume} {45}},\ \bibinfo {pages} {11415}
  (\bibinfo {year} {1992})}\BibitemShut {NoStop}%
\bibitem [{\citenamefont {Gurlu}\ \emph {et~al.}(2004)\citenamefont {Gurlu},
  \citenamefont {Zandvliet}, \citenamefont {Poelsema}, \citenamefont {Dag},\
  and\ \citenamefont {Ciraci}}]{Gurlu_PRB_2004}%
  \BibitemOpen
  \bibfield  {author} {\bibinfo {author} {\bibfnamefont {O.}~\bibnamefont
  {Gurlu}}, \bibinfo {author} {\bibfnamefont {H.~J.~W.}\ \bibnamefont
  {Zandvliet}}, \bibinfo {author} {\bibfnamefont {B.}~\bibnamefont {Poelsema}},
  \bibinfo {author} {\bibfnamefont {S.}~\bibnamefont {Dag}}, \ and\ \bibinfo
  {author} {\bibfnamefont {S.}~\bibnamefont {Ciraci}},\ }\Doi
  {10.1103/PhysRevB.70.085312} {\bibfield  {journal} {\bibinfo  {journal}
  {Phys. Rev. B},\ }\textbf {\bibinfo {volume} {70}},\ \bibinfo {pages}
  {085312} (\bibinfo {year} {2004})}\BibitemShut {NoStop}%
\bibitem [{\citenamefont {Smith}\ \emph {et~al.}(1989)\citenamefont {Smith},
  \citenamefont {Luo}, \citenamefont {Hashimoto}, \citenamefont {Gibson},\ and\
  \citenamefont {Lewis}}]{Smith_JVSTA_89}%
  \BibitemOpen
  \bibfield  {author} {\bibinfo {author} {\bibfnamefont {G.~A.}\ \bibnamefont
  {Smith}}, \bibinfo {author} {\bibfnamefont {L.}~\bibnamefont {Luo}}, \bibinfo
  {author} {\bibfnamefont {S.}~\bibnamefont {Hashimoto}}, \bibinfo {author}
  {\bibfnamefont {W.~M.}\ \bibnamefont {Gibson}}, \ and\ \bibinfo {author}
  {\bibfnamefont {N.}~\bibnamefont {Lewis}},\ }\Doi {10.1116/1.576080}
  {\bibfield  {journal} {\bibinfo  {journal} {J. Vac. Sci.Tech. A},\ }\textbf
  {\bibinfo {volume} {7}},\ \bibinfo {pages} {1475} (\bibinfo {year}
  {1989})}\BibitemShut {NoStop}%
\bibitem [{\citenamefont {Garleff}\ \emph {et~al.}(2004)\citenamefont
  {Garleff}, \citenamefont {Wenderoth}, \citenamefont {Sauthoff}, \citenamefont
  {Ulbrich},\ and\ \citenamefont {Rohlfing}}]{Rohlfing_PRB_04}%
  \BibitemOpen
  \bibfield  {author} {\bibinfo {author} {\bibfnamefont {J.~K.}\ \bibnamefont
  {Garleff}}, \bibinfo {author} {\bibfnamefont {M.}~\bibnamefont {Wenderoth}},
  \bibinfo {author} {\bibfnamefont {K.}~\bibnamefont {Sauthoff}}, \bibinfo
  {author} {\bibfnamefont {R.~G.}\ \bibnamefont {Ulbrich}}, \ and\ \bibinfo
  {author} {\bibfnamefont {M.}~\bibnamefont {Rohlfing}},\ }\Doi
  {10.1103/PhysRevB.70.245424} {\bibfield  {journal} {\bibinfo  {journal}
  {Phys. Rev. B},\ }\textbf {\bibinfo {volume} {70}},\ \bibinfo {pages}
  {245424} (\bibinfo {year} {2004})}\BibitemShut {NoStop}%
\end{thebibliography}
\end{document}